\documentclass[a4paper,11pt]{article}
\pdfoutput=1 

\usepackage{jheppub} 

\usepackage[T1]{fontenc} 

\newcommand{\beq}{\begin{equation}}
\newcommand{\eeq}{\end{equation}}
\newcommand{\bea}{\begin{eqnarray}}
\newcommand{\eea}{\end{eqnarray}}
\newcommand{\vep}{\varepsilon}
\newcommand{\vH}{\upsilon_H}

\newcommand{\nn}{\nonumber}
\newcommand{\mX}{M_\chi}

\newcommand{\meFN}{m_{\rm e}^{\rm FN}}
\newcommand{\muFN}{m_{\mu}^{\rm FN}}
\newcommand{\merad}{m_{\rm e}^{\rm RAD}}
\newcommand{\murad}{m_{\mu}^{\rm RAD}}
\newcommand{\e}{{\rm e}}

\newcommand{\fref}[1]{Figure~\ref{#1}} 
\newcommand{\eref}[1]{Eq.(\ref{#1})}

\newcommand{\aref}[1]{Appendix~\ref{#1}}
\newcommand{\sref}[1]{Section~\ref{#1}}

\title{\boldmath Muon and electron $g-2$ and   lepton masses in flavor models }

\author[a]{Lorenzo Calibbi}
\author[b]{,\,M.L.~L\'{o}pez-Ib\'{a}\~{n}ez}
\author[c]{,\,Aurora Melis}
\author[c]{and Oscar Vives}

\preprint{IFIC/20-09, FTUV-20-0314}

\affiliation[a]{School of Physics, Nankai University,\\Tianjin 300071, China}
\affiliation[b]{CAS Key Laboratory of Theoretical Physics, Institute of Theoretical Physics, Chinese Academy of Sciences, \\Beijing 100190, China}
\affiliation[c]{Departament de F\'{i}sica T\`{e}orica, Universitat de Val\`{e}ncia \& IFIC, Universitat de Val\`{e}ncia \& CSIC, \\Dr.~Moliner 50, E-46100 Burjassot (Val\`{e}ncia), Spain}

\emailAdd{calibbi@nankai.edu.cn}
\emailAdd{maloi2@uv.es}
\emailAdd{aurora.melis@uv.es}
\emailAdd{oscar.vives@uv.es}

\abstract{The stringent experimental bound on $\mu \rightarrow \e \gamma$ is compatible with a simultaneous and sizable new physics contribution to the electron and muon anomalous magnetic moments $(g-2)_\ell$ ($\ell=\e,\,\mu$), only if we assume a non-trivial flavor structure of the dipole operator coefficients. We propose a mechanism in which the realization of the $(g-2)_\ell$ correction is manifestly related to the mass generation through a flavor symmetry. A radiative flavon correction to the fermion mass gives a contribution to the anomalous magnetic moment. In this framework, we introduce a chiral enhancement from a non-trivial $\mathcal{O}(1)$ quartic coupling of the scalar potential. We show that the muon and electron anomalies can be simultaneously explained in a vast region of the parameter space with predicted vector-like mediators of masses as large as $M_\chi\in [0.6,2.5]$~TeV.
}

\begin{document} 
\maketitle
\flushbottom

\section{Introduction}
\label{sec:intro}
Despite the lack of direct signals for new physics from the high-energy collision data collected by the LHC experiments, 
we have a number of solid arguments, both theoretical and observational, that call for extensions of the Standard Model (SM).
The most convincing of those\,---\,related to the origin of neutrino masses, dark matter, baryon asymmetry etc.\,---\,do not necessarily point to new particles at scales accessible at colliders in the foreseeable future.
However, recent years have been also witnessing the arising of several hints for non-standard phenomena from precision observables involving lepton flavors. 
Signs of departure from the universality of leptonic couplings predicted by the SM in semi-leptonic decays of $B$ mesons have been reported by LHCb and B-factories experiments both in neutral- and charged-current processes\,---\,for recent reviews see \cite{Albrecht:2018vsa,Li:2018lxi,Bifani:2018zmi}. If confirmed by future data, these discrepancies would certainly require low-scale new physics coupling with different strength to different lepton families. 
Another discrepancy that would point to an analogous conclusion is related to the anomalous magnetic moment of the muon, $(g-2)_\mu$. 
The experimental measurements of $(g-2)_\mu$ have been in tension with the increasingly accurate theoretical calculations within the SM for about 20 years. The discrepancy currently amounts to about $3.5\,\sigma$ \cite{Bennett:2006fi,Davier:2010nc,Davier:2017zfy,Keshavarzi:2018mgv,Blum:2018mom,Campanario:2019mjh,Davier:2019can}.\footnote{See, however, the very recent lattice result of the leading order hadronic vacuum polarization \cite{Borsanyi:2020mff}, which, contrary to previous results, could reduce this discrepancy. On the other hand, even if the anomaly is accounted for by the hadronic vacuum polarization, this would reflect in a deterioration of the EW fit and the arising of tensions of comparable significance in other observables \cite{Passera:2008jk,Crivellin:2020zul}.} The situation may be clarified\,---\,and the case for new physics possibly reinforced\,---\,by the upcoming results of the new Muon g-2 experiment at FNAL \cite{Grange:2015fou}.
It is well known that new particles coupling to muons can easily account for the $(g-2)_\mu$ provided that their mass are few TeV at most\,---\,for a recent review see \cite{Lindner:2016bgg}. This makes the new physics possibly required by the $(g-2)_\mu$ anomaly an ideal target for direct searches at LHC experiments, which in fact have already reached the sensitivity so to exclude substantial portions of the parameter space of typical models~\cite{Moroi:1995yh,Martin:2001st,Stockinger:2006zn,Endo:2013bba,Lindner:2016bgg,Endo:2020mqz}. 

Interestingly, a 2$\sigma$ tension between theory and experiments has been recently reported also for the electron $g-2$. 
The $(g-2)_\e$ has been determined both experimentally and theoretically to such an outstanding precision, that matching the SM prediction to the measurement has been used for many years as the most precise way to evaluate the fine-structure constant $\alpha$. However, in presence of an alternative and sufficiently precise measurement of $\alpha$, one can employ $(g-2)_\e$ as a test for new physics too \cite{Giudice:2012ms}. This has become possible in recent years 
and the most precise result, obtained by employing matter-wave interferometry with cesium-133 atoms \cite{Parker:2018vye}, highlighted the discrepancy for  $(g-2)_\e$ mentioned above. Expressed in terms of $a_\ell \equiv(g-2)_\ell/2$, the present situation can be summarized as follows:
\begin{align}
\Delta a_\e^{\rm exp}\equiv &~a_\e^{\rm exp} - a_\e^{\rm SM} = -(8.8\pm 3.6) \times 10^{-13}, \label{eq:dae} \\
    \Delta a_\mu^{\rm exp} \equiv &~a_\mu^{\rm exp} - a_\mu^{\rm SM} = (2.7 \pm 0.7) \times 10^{-9}. \label{eq:dam}
\end{align}

It is very tempting to speculate about a simultaneous new-physics origin of the results above, outlining the same mechanism or, at least, a single model able to explain both discrepancies. In fact, this has been recently attempted in a number of works \cite{Davoudiasl:2018fbb, Crivellin:2018qmi,Liu:2018xkx,Han:2018znu,Endo:2019bcj,Abdullah:2019ofw,Bauer:2019gfk,Badziak:2019gaf,CarcamoHernandez:2019ydc,Hiller:2019mou,Cornella:2019uxs,Endo:2020mev,Jana:2020pxx}. Although a common explanation has been shown to be possible, the model building task has proved non-trivial. 
First, as Eqs.(\ref{eq:dae},\,\ref{eq:dam}) show, the new-physics contributions
need to be positive for $\Delta a_\mu$ and negative for $\Delta a_\e$.
Secondly, the absolute magnitude of $\Delta a_\mu$ and $\Delta a_\e$ do not match the naive scaling $\Delta a_\mu/\Delta a_\e \sim m_\mu^2/m_\e^2$ \cite{Giudice:2012ms} expected in models where the chirality flip of the lepton field in the dipole operator is provided by the lepton Yukawa coupling itself\,---\,see discussion below. In fact, such a scaling would result in an absolute value for $\Delta a_\e$ way too suppressed compared to the experimental range in \eref{eq:dae}.
New physics giving a chirally-enhanced contribution\,---\,i.e.~featuring the chirality flip inside the loop\,---\,at least to $\Delta a_\e$ is thus required in order to account for Eqs.(\ref{eq:dae},\,\ref{eq:dam}) simultaneously.
The third and perhaps most important challenge model building has to face concerns the tight experimental
limits on lepton-flavor-violating (LFV) processes\,---\,see e.g.~\cite{Calibbi:2017uvl} for a recent review\,---\,in particular $\mu\to \e\gamma$. It is clear that any new physics contributing to both the electron and the muon dipole moment will in general induce the corresponding $\mu-\e$ dipole transition. 

We can quantify the above difficulties as follows.
In an effective Lagrangian approach, non-standard effects to the leptonic observables of interest ($\Delta a_\ell$, $\mu\to \e\gamma$, EDMs, etc.) arise via the dipole operators:
\beq
    {\cal L} \:\supset\: \frac{e\, {m^{\rm exp}_{\ell}}}{8\, \pi^2}\, C_{\ell \ell^\prime}\left(\bar\ell \sigma_{\mu \nu} P_R \ell^\prime\right)\, F^{\mu\nu} \:+\: {\rm h.c.} \quad\quad \ell,\ell^\prime =\e,\mu,\tau.
\label{eq:L-dipole}
\eeq
This effective Lagrangian constitutes a model-independent description of the new-physics effects we are interested in, so long as the new-physics scale is much larger than the energy scale associated to our observables, i.e.~the lepton masses.
In terms of the above Wilson coefficients\,---\,that in our convention have mass dimension $\rm GeV^{-2}$\,---\,the new-physics contribution to the $\Delta a_\ell$ reads:
\beq
    \Delta a_\ell = \frac{{{m^{\rm exp\,}_{\ell}}^2}}{(2 \pi)^2}\, \mbox{\rm Re} (C_{\ell \ell}).
\eeq
In order to fit the experimental results\,---\,for illustration we focus here on the central values in  Eqs.(\ref{eq:dae},\:\ref{eq:dam})\,---\,the dipole coefficients need to attain the following numerical values:
 \bea
    \mbox{\rm Re}(C_{\e\e})       & \approx & - \,7 \times 10^{-5}~~ \mbox{\rm GeV}^{-2}, \label{eq:Cee} \\
    \mbox{\rm Re}(C_{\mu\mu})   & \approx & \quad 5 \times 10^{-6}~ ~\mbox{\rm GeV}^{-2}. \label{eq:Cmm}
\eea
The flavor-changing couplings instead contribute to LFV processes, in particular to the radiative decays:
\begin{align}
\frac{{\rm BR}(\ell\to {\ell^\prime} \gamma)}{{\rm BR}(\ell\to \ell^\prime \nu \bar{\nu}^\prime)} = \frac{3\alpha}{\pi G_F^2} \left( 
|C_{\ell \ell^\prime}|^2 + |C_{\ell^\prime \ell}|^2\right),
\end{align}
where the coefficients $C_{\ell \ell^\prime}$ are defined in the basis where the lepton Yukawa matrix $Y_\ell$ is diagonal.
The experimental bound ${\rm BR}(\mu\to \e \gamma) < 4.2\times 10^{-13}$ \cite{TheMEG:2016wtm} then translates into the following constraint:
\bea \label{eq:Cem}
    |C_{\e\mu}|,~| C_{\mu\e}| \:\lesssim\:  10^{-10}~~ \mbox{\rm GeV}^{-2}.
\eea

Notice that defining the coefficients in \eref{eq:L-dipole}
we have factored out the dependence on the lepton masses. Hence, in models where the chirality flip of the lepton fields required by gauge invariance in \eref{eq:L-dipole} is due to a lepton mass insertion, the coefficients
$C_{\e\e}$ and $C_{\mu\mu}$ should be of the same order $1/\Lambda^2$, where $\Lambda$ is the scale of new physics, with no further chirality suppression. Nevertheless, Eqs.(\ref{eq:Cee},\:\ref{eq:Cmm})
tell us that this would result in a contribution to the electron magnetic moment a factor of $15$ too small.
If, on the other hand, the chirality flip in \eref{eq:L-dipole} is due to the insertion of a Higgs vev inside the loop, one expects an enhancement
of the order $C_{\ell\ell}\sim y_\chi/y_\ell$ where $y_\chi$ is the coupling of the new fields to the Higgs and $y_\ell$ is the lepton Yukawa\,---\,see e.g.~the discussion in \cite{Calibbi:2018rzv}. If the same coupling $y_\chi$ enters the diagrams for the electron and the muon dipole moment, one would then obtain $\Delta a_\mu/\Delta a_\e \sim m_\mu/m_\e$. Again this is not compatible with the observed ranges of Eqs.(\ref{eq:dae},\:\ref{eq:dam}): besides the sign, in this case the contribution to the electron $g-2$ would result about a factor 15 too large.

From this discussion, it is clear that suitable new physics contributions should be flavor-dependent and rather sizable without disturbing the small values of the electron and muon masses\,---\,any loop contributing to dipole operators would generate a radiative contribution to lepton masses as well\,---\,and without being in conflict with LFV constraints. In fact, Eqs.(\ref{eq:Cee},\:\ref{eq:Cmm}) and \eref{eq:Cem} show that a simultaneous explanation of the two anomalies requires a relative suppression of the LFV coefficients by more than five orders of magnitude. In other words, the matrix $C_{\ell\ell^\prime}$ and the lepton Yukawa matrix have to be almost aligned in flavor space, to such extent that the relative misalignment angle can not exceed $\mathcal{O}(10^{-6})$. 
A priori there is no reason why generic new physics responsible of non-standard $g-2$ of leptons should have a flavor structure so perfectly aligned to the SM lepton mass matrix, unless of course the two sectors share a common origin. Hence we find it natural to investigate the possibility of a combined explanation of the electron and muon $g-2$ within a model of flavor, i.e.~directly arising from the same dynamics behind the observed lepton masses.
Our idea is to focus on flavor models {\it \`a la} Froggatt-Nielsen \cite{Froggatt:1978nt,Leurer:1992wg,Leurer:1993gy} and calculate the contribution to the lepton $g-2$ of the flavons and the mediator fields that generate the charged-lepton masses.

The rest of the paper is organized as follows. In \sref{sec:theory} we highlight the general idea and the fundamental ingredients to obtain successful lepton masses and $g-2$ from a flavor model. Section~\ref{sec:toy} shows how this is realized in a toy model example. In Section~\ref{sec:Pheno} we discuss the phenomenology of flavons and mediators and we conclude in Section~\ref{sec:Conclusions}.

\section{General Idea}
\label{sec:theory}

As discussed in the introduction, the new contributions to the anomalous magnetic moment must be flavor dependent, but with a different flavor dependence from the SM Yukawa couplings.\footnote{However, for an exception see Ref.~\cite{Hiller:2019mou}}  Although, in principle, it would be possible to assign an {\it ad hoc} flavor structure, both to the magnetic moments and to the Yukawas, it is more satisfactory to try to explain these observables in terms of a new symmetry in flavor space.
Indeed, flavor symmetries {\it \`a la} Froggatt-Nielsen (FN) have been used for a long time to understand the complex structure of Yukawa couplings. In this framework, it looks completely natural to use the same mechanism to explain the new structures of dipole operators.\footnote{During the completion of this work, an article appeared~\cite{Haba:2020gkr}, that also proposes a possible connection between anomalous magnetic moments and a $U(1)$ flavor symmetry, although in the context of a multi-Higgs doublet model rather than FN models.}
Fermion masses and anomalous magnetic moments, both chirality changing operators, are intimately connected. Any radiative correction to the fermion masses gives a contribution to the anomalous magnetic moment if we attach a photon to one of the internal lines. However, the FN contributions to the Yukawas usually considered are tree-level diagrams while we necessarily need a loop to generate the dipole operators. In any case, loop corrections to the tree-level diagrams are always present and, as we will see below, under certain conditions they can be sizeable with respect to the tree-level diagrams. 

\begin{figure}[t]
  \centering
  \includegraphics[scale=0.65]{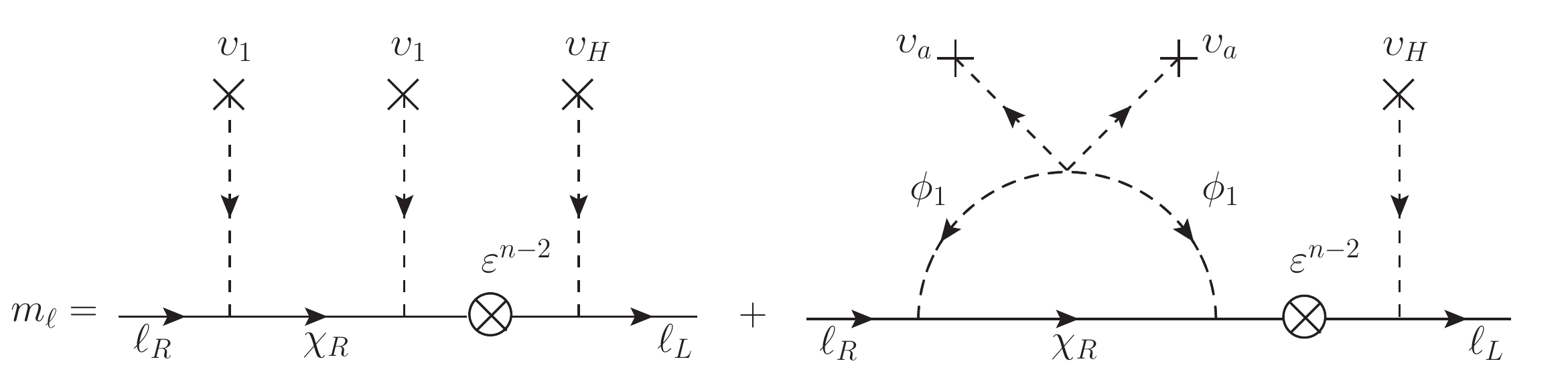}
  \caption{\label{fig:Masses} Froggatt Nielsen (Left) and Radiative (Right) lepton mass.}\vspace{6mm}
  \includegraphics[scale=0.65]{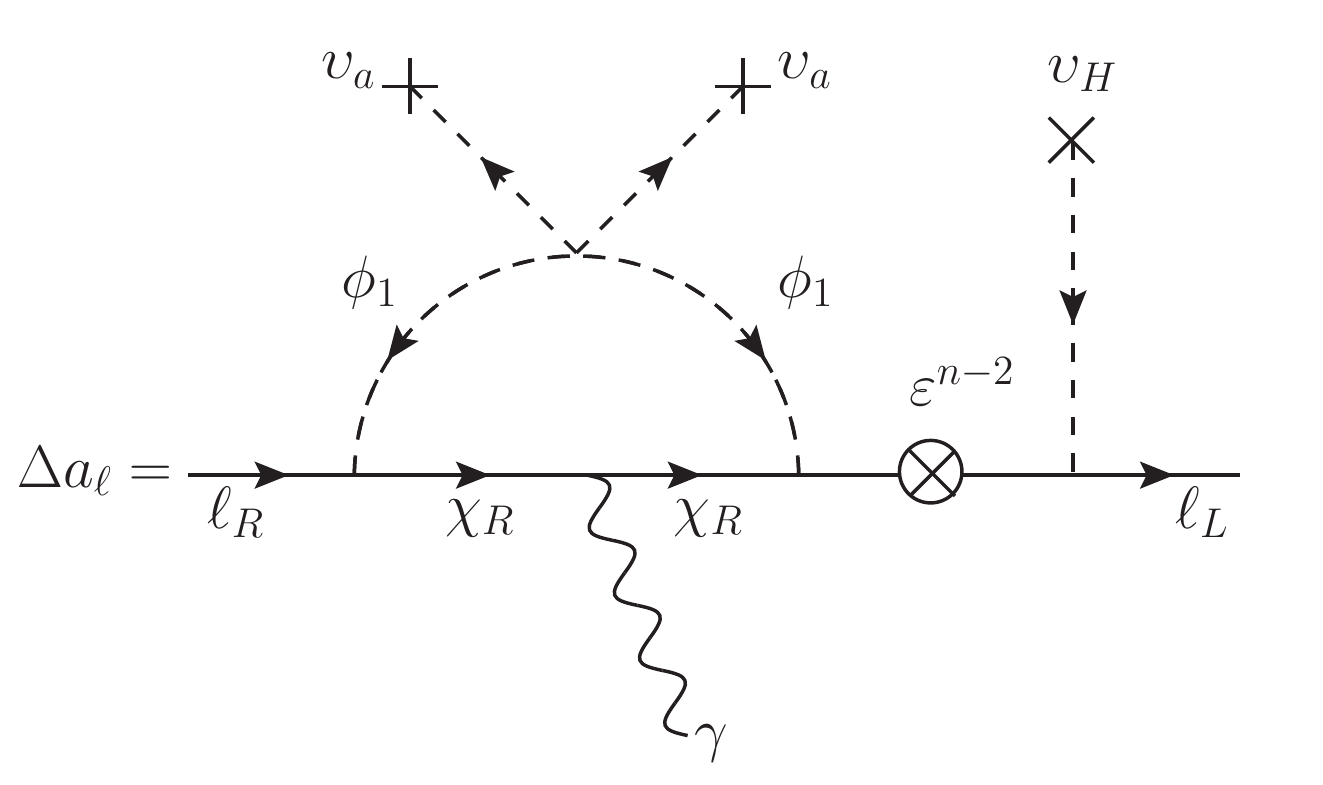}\caption{\label{fig:(g-2)m} Flavon contribution to $(g-2)_\ell$}
\end{figure}

Yukawa couplings are accounted for as powers of a dimensionless ratio $\upsilon/M \leq 1$, with $\upsilon$ a scalar vacuum expectation value, singlet under the SM symmetries, and $M$ the mass of a heavy vector-like mediator with the SM quantum numbers. These contributions are obtained from tree-level diagrams as shown in Figure~\ref{fig:Masses}. 
Nevertheless, the radiative corrections to this diagram can be large. In particular, we could consider loops involving the flavons with small vevs, so that we could ``replace'' two small vevs by an ${\cal O} (1)$ loop function. Obviously, this is not so easy, as the flavons carry a flavor charge and they must break the symmetry to connect the low energy fermionic fields and thus the loop must also break the symmetry by the same amount. 
This could be done through the flavon vev itself. However, as we will see in the following, the above mentioned enhancement can  be achieved only if a larger vev of a different flavon field is inserted, being the size of this vev not fixed if this field does not couple directly to the fermions. In this way, it is possible to partially compensate the loop suppression and make this loop contribution, with the correct symmetry-breaking properties,  comparable to the tree-level FN diagram.

Now, it is clear that this loop diagram generating a loop correction to the Yukawa would be the same as the diagram generating the dipole coefficients simply adding a photon, see \fref{fig:(g-2)m}.  In general, we expect that the anomalous magnetic moment $a_\ell = C m_\ell^2/M^2$ \cite{Czarnecki:2001pv} with $C$ a loop factor if the fermion mass is present at tree-level or $C\sim {\cal O}(1)$ if the mass is generated at loop level \cite{Okada:2013iba}. In our flavor symmetry models, we could have radiative corrections to the mass similar to the tree-level contribution which implies that a large contribution to $a_\ell$, with $C\sim {\cal O}(1)$, can be expected. Moreover, the measured discrepancies in the muon and electron magnetic moments, which do not follow this quadratic scaling with the fermion mass, can also be explained with flavor models where additional flavor dependence  can enter naturally the magnetic moment. The main problem of this construction, as discussed in the introduction, is to suppress off-diagonal LFV dipole operators which requires some non-trivial model building.

 On the other hand, in flavor symmetry models, the dimensionless Yukawa couplings depend only on ratios $\upsilon/M$ and therefore can not fix the scale of symmetry breaking or the mediator masses. Anomalous magnetic moments are dimension 6 operators, and then the contributions to $a_\ell$ are suppressed, compared to the radiative contribution to the mass, by the heaviest mass in the loop, i.e.~in our flavor models, the mediator mass, $M_\chi^2$, or the flavon mass, $M_\phi^2$.  
 Therefore, this implies that anomalous magnetic moments could provide a hint on the scale of flavor symmetry breaking if the measured discrepancies are due to these flavon contributions. 
 
 At this point, we would like to emphasize that the relation between anomalous magnetic moments and radiative corrections to the masses is true for a generic model. In particular, models with a chiral enhancement in the lepton anomalous magnetic moments can also have large corrections to the tree-level lepton masses.    This is what happens, for instance, in the MSSM with large $\tan \beta$ or in models with leptoquarks (LQs) where the chirality flip can be given by a  quark mass, e.g.~$m_t$, instead of $m_\mu$ or $m_\e$. 
In fact,  the required contributions to the anomalous magnetic moments generically imply a large correction to the mass, which is usually not taken into account in the literature. For instance, models with multi-TeV chirality-flipping vector-like leptons or LQs that can explain $(g-2)_\mu$, as is the case in Refs.~\cite{Bauer:2015knc,ColuccioLeskow:2016dox,Crivellin:2017dsk,Calibbi:2018rzv,Dorsner:2019itg,Calibbi:2019bay,Crivellin:2019dwb,Altmannshofer:2020ywf,Bigaran:2020jil}, could give a sizable correction to the mass. Assuming the loop functions in the radiative mass and anomalous magnetic moments to be of the same order, we can estimate $m_\mu^{\rm RAD} \sim  \Delta a_\mu ~M_\chi^2 /2 m_\mu \sim 0.05 ~(M_\chi/2\,{\rm TeV})^2$~GeV, with $M_\chi$ the leptoquark mass. This large contribution could cancel against a tree level mass contribution with some degree of tuning, but radiative corrections to the mass should be taken into account in these analysis.

{Notice that, in this work, we concentrate on the charged-lepton sector and we do not discuss neutrino mixings. The observed neutrino mixings can always be accommodated with the help of the right-handed neutrino mass matrices in a type-I seesaw mechanism, possibly with additional breaking of the flavor symmetry.}
In the following, we apply these general ideas to explain the measured discrepancies $\Delta a_\e$ and $\Delta a_\mu$ in models of flavor symmetries. For this, we will construct an explicit example of this mechanism. 
 
\subsection[Lepton masses and $g-2$ contribution]{Lepton masses and \texorpdfstring{\boldmath $g-2$}{Lg} contribution}
\label{subsec:MassesFN}
Assuming a minimal set of fields and couplings, the Yukawa-like interactions responsible for the masses in a FN framework can be schematically written as:
\bea \label{eq:secILy}
    {\cal L}_{Y} & = & g_\ell \left(\, \overline\chi_R\, \ell_R\, \phi_1 \;+\; 
                    \dots \;+\; \overline\ell_L\,\chi_R\, H \,\right) \;+\; {\rm h.c.},
\eea
with $\chi_R$ a heavy vector-like mediator with the quantum number of a right-handed lepton $\ell_R$,\footnote{Obviously one could also consider mediators carrying the quantum numbers of left-handed leptons, or a combination of right-handed and left-handed mediators with the Higgs not coupling directly to the light chiral fields. For a detailed discussion of the messenger sector of FN models see \cite{Calibbi:2012yj,Calibbi:2012at,Das:2016czs,Lopez-Ibanez:2017xxw}.}
$\phi_1$ a flavon field carrying non-zero flavor charge,
and $g_\ell$ a generic ${\cal O}(1)$ coupling
that, for illustration purposes, we took to be the same for all interactions. As we will see below, our results not depending on this choice.
Then, the minimal potential should contain the following couplings:
\beq
\label{eq:Vphi}
    V(\phi) \:=\: \sum_{i}\, -\mu_i^2\, (\phi_i^\dagger \phi_i) + \lambda_i  (\phi_i^\dagger \phi_i)^2 \:+\: \frac{1}{2}\sum_{i\neq j}\, \lambda_{ij}\, (\phi_i^\dagger\phi_i) (\phi_j^\dagger\phi_j) \:+\: \bigg[\lambda\, (\phi_a^\dagger \phi_1)^2 \:+\: {\rm h.c.} \bigg]\,,
\eeq
where the indices $i,j=1,a,\dots$ go through all the flavons present in the model. We have introduced $\phi_a$ as a general complex scalar field that does not couple directly to leptons. Other quartic terms of the kind $(\phi_i^\dagger \phi_j)^2$ could also be present in Eq.\eqref{eq:Vphi} provided they respect the flavor symmetry. 
The interactions in Eq.\eqref{eq:secILy} induce a mass term for the charged leptons through the processes depicted in Figure~\ref{fig:Masses}.
For a diagram with $n$ flavon insertions, the effective mass is:
\bea 
\label{eq:MassFN}
    m_\ell^{\rm FN}   \:=\: g^{n+1}_\ell\, \frac{\vH}{\sqrt{2}}\,\vep^{n-2}\,\left(\frac{\upsilon_1}{M_\chi}\right)^{2}\,,
\eea
where $\upsilon_1 \equiv \langle \phi_1 \rangle$, $M_\chi$ is the heavy mediator mass and $\vep \equiv \upsilon / M_\chi$ stands for possible additional insertions of the same $\phi_1$ as well as of other flavons generally present in a complete flavor model. Depending on the number of different flavons and vertices in the lagrangian, we have to take into account possible degeneracy coefficients, which count for the possible ways of inserting each flavon. However, they can always be absorbed into the $g_\ell$ coupling.
 In Figure~\ref{fig:Masses}, we show that, together with the FN-diagram, the last vertex of Eq.\eqref{eq:Vphi} also induces a radiative mass term $m_\ell^{\rm RAD}$. The computation of the diagram gives:
\beq \label{eq:RadMass}
    m_\ell^{\rm RAD} \:=\: g^{n+1}_\ell\, \frac{\vH}{\sqrt{2}}\,\varepsilon^{n-2}\,\left(\frac{\upsilon_a}{M_\chi}\right)^{2}
    \,\frac{\lambda}{16 \pi^2}\,I_m^\times(x_\phi),
\eeq
and the following loop function 
\beq \label{eq:MassInt}
    I_m^{\times}(x_\phi) \:=\:  \frac{1+2\log 
    x_\phi -x_\phi^2 }{(1-x_\phi^2)^2} < 0\,,
\eeq
with $x_\phi=\mu_{\phi_1}/M_\chi$, being $\mu_{\phi_1}^2$ the bilinear coupling in the scalar potential before symmetry breaking. We must remark here that $\phi_1$ is a complex scalar and the FN-operator involves $\phi_1^2$, therefore a bilinear coupling, $\mu_i^2$, can not close the loop in Figure~\ref{fig:Masses} and we must take a quartic couplings with two vevs breaking the flavor symmetry.
Comparing Eqs.~(\ref{eq:MassFN}) and (\ref{eq:RadMass}), they differ for the loop factor $\lambda\,I_m^\times /(16\,\pi^2)$ and the replacement $\upsilon_1\rightarrow \upsilon_a$. The contribution $m_\ell^{\rm RAD}$ is comparable with $m_\ell^{\rm FN}$ if $(\upsilon_a/\upsilon_1)^2$ is big enough to compensate for the suppression of the loop factor. Note that, if more than two insertions of $\phi_1$ are present, we have to take into account the alternative ways of closing the loop. As a consequence, typically there is a mismatch between the degeneracy coefficients of the FN and RAD diagrams that can not be reabsorbed. The function in Eq.(\ref{eq:MassInt}) is defined negative; this means that as long as  $\lambda >0$, the two diagrams in Figure~\ref{fig:Masses} interfere destructively among each other.

What we want to emphasize is that the same processes which generates the radiative contribution to the lepton masses induces a correction to the anomalous magnetic moment coupling a photon to the loop.
It contributes to the anomalous magnetic moment as:
\bea \label{eq:(g-2)}
    \Delta a_\ell & = & g_\ell^{n+1} \frac{\vH}{\sqrt{2}}\,\frac{m^{\rm exp}_\ell}{\mX^2}\,
   \varepsilon^{n-2}\,      \left(\frac{\upsilon_a}{M_\chi}\right)^2
                        \,\frac{\lambda}{8\, \pi^2}\,I_{\Delta a}^\times(x_\phi).
\eea
where the loop function is given by
\bea \label{eq:(g-2)Int}
    I_{\Delta a}^\times(x_\phi) \:=\: -\frac{1 \,+\, 4x_\phi^2(1\,+\,2\log x_\phi)\, -\, x_\phi^4(5-4\log x_\phi)}{2(1-x_\phi^2)^4} <0 \,.\eea

\begin{figure}[t!]
  \centering
        \includegraphics[scale=0.266]{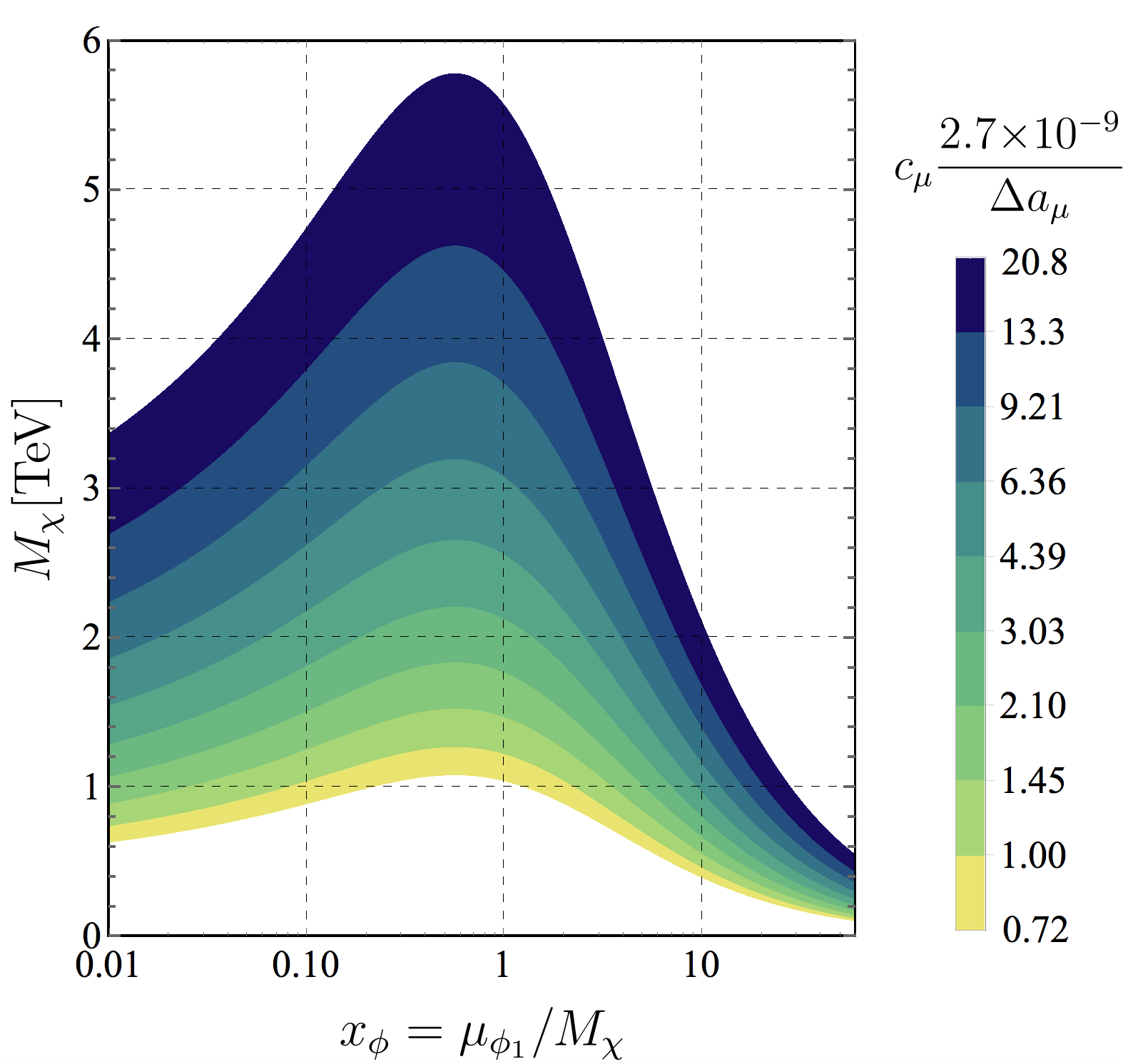}\hspace{-1mm}
        \includegraphics[scale=0.275]{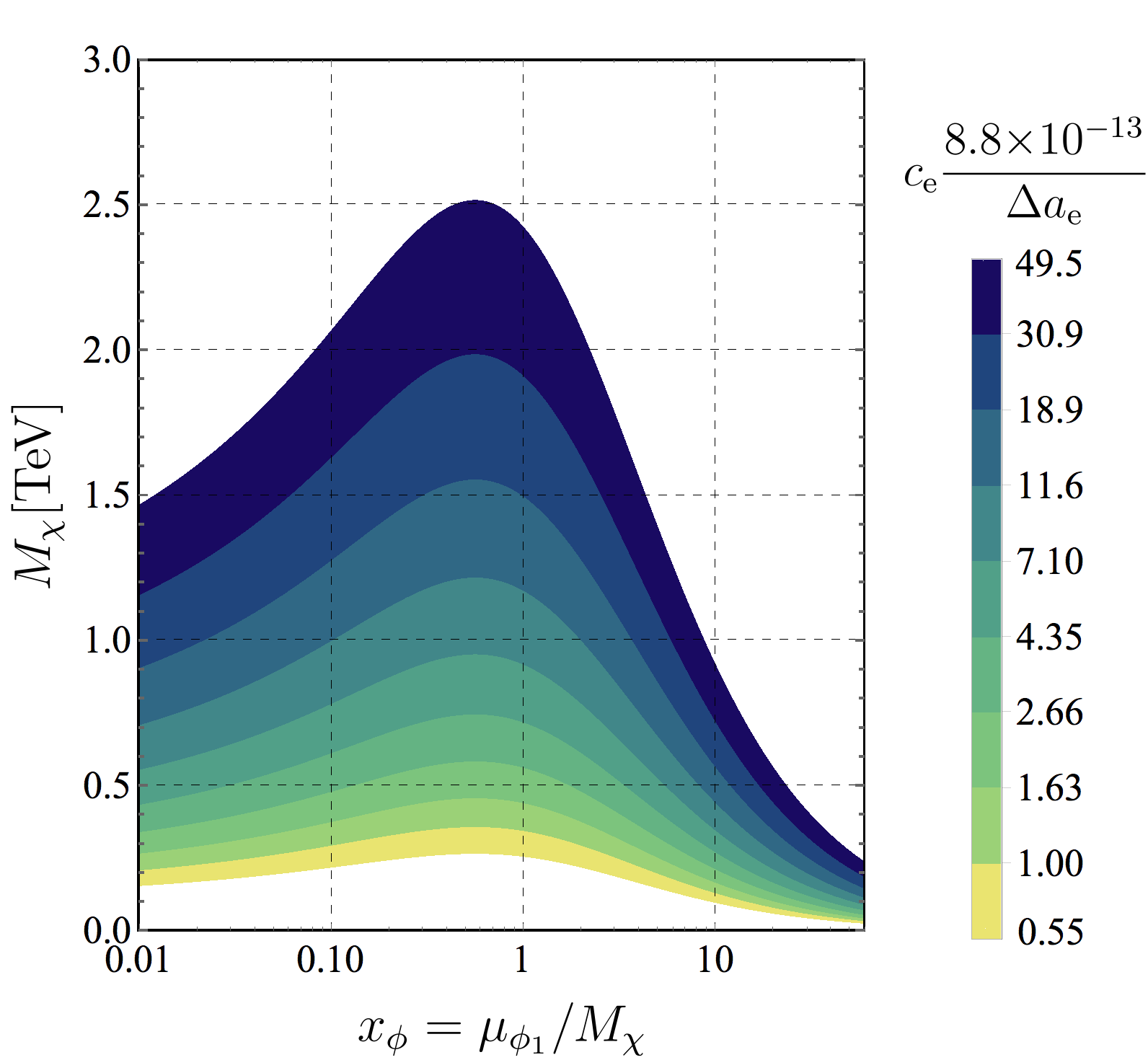}
        \caption{\label{fig:MXxphi} Mediator mass $M_\chi$ as function of $x_\phi$ for the muon case (Left panel) and the electron case (Right panel). We have used Eq.(\ref{eq:Da_vs_mRAD}) imposing $\Delta a^{\rm exp}_{\ell} - 2 \sigma_{\ell} < \Delta  a_{\ell} < \Delta a^{\rm exp}_{\ell} + 2 \sigma_{\ell}$ and $c_\ell \in [1,10]$. For a given value of $\Delta a_\ell$, we can read the required loop factor, $c_\ell$, from the bar legend.} \vspace{5mm}  \includegraphics[scale=0.298]{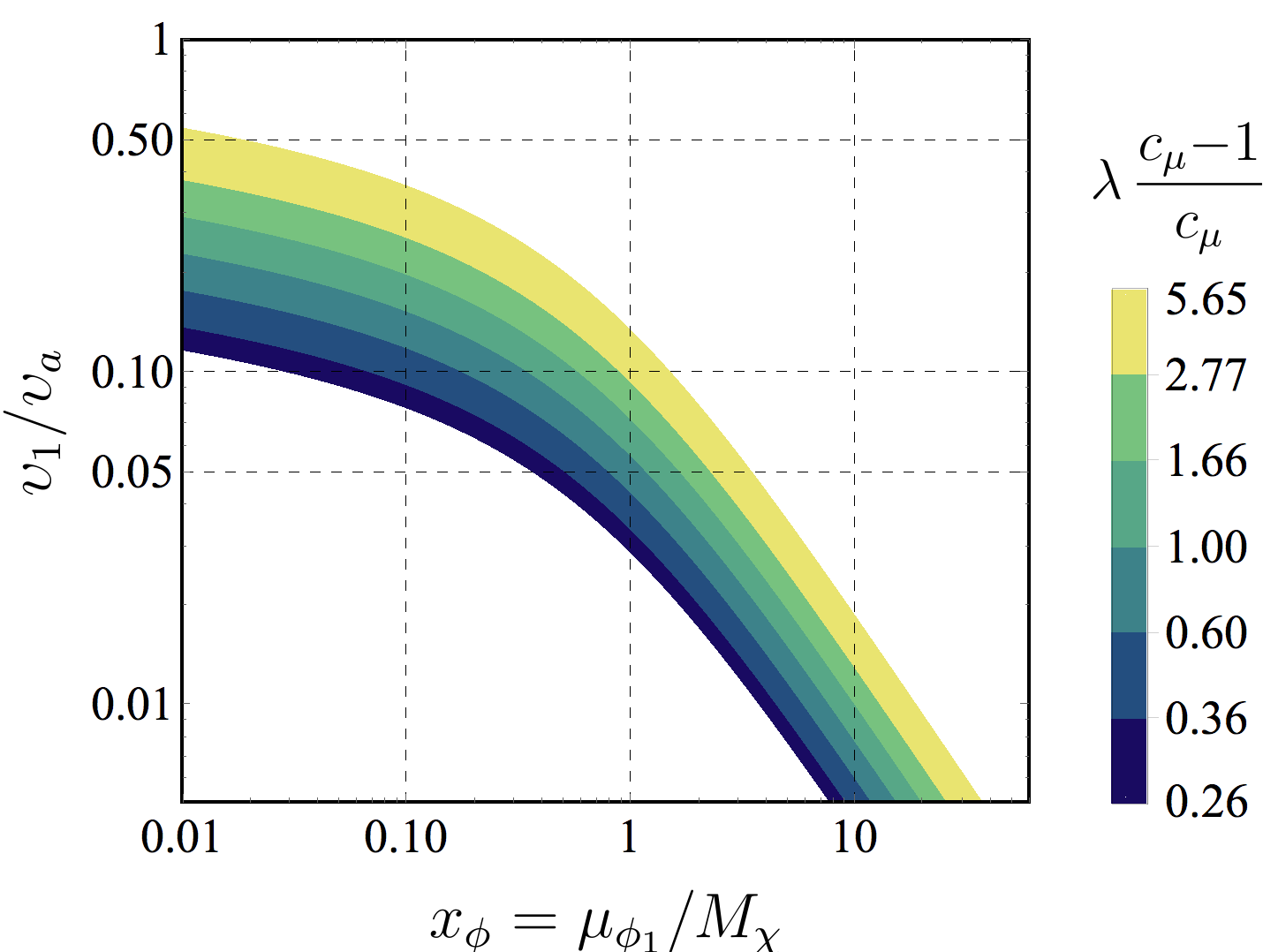}
        \includegraphics[scale=0.296]{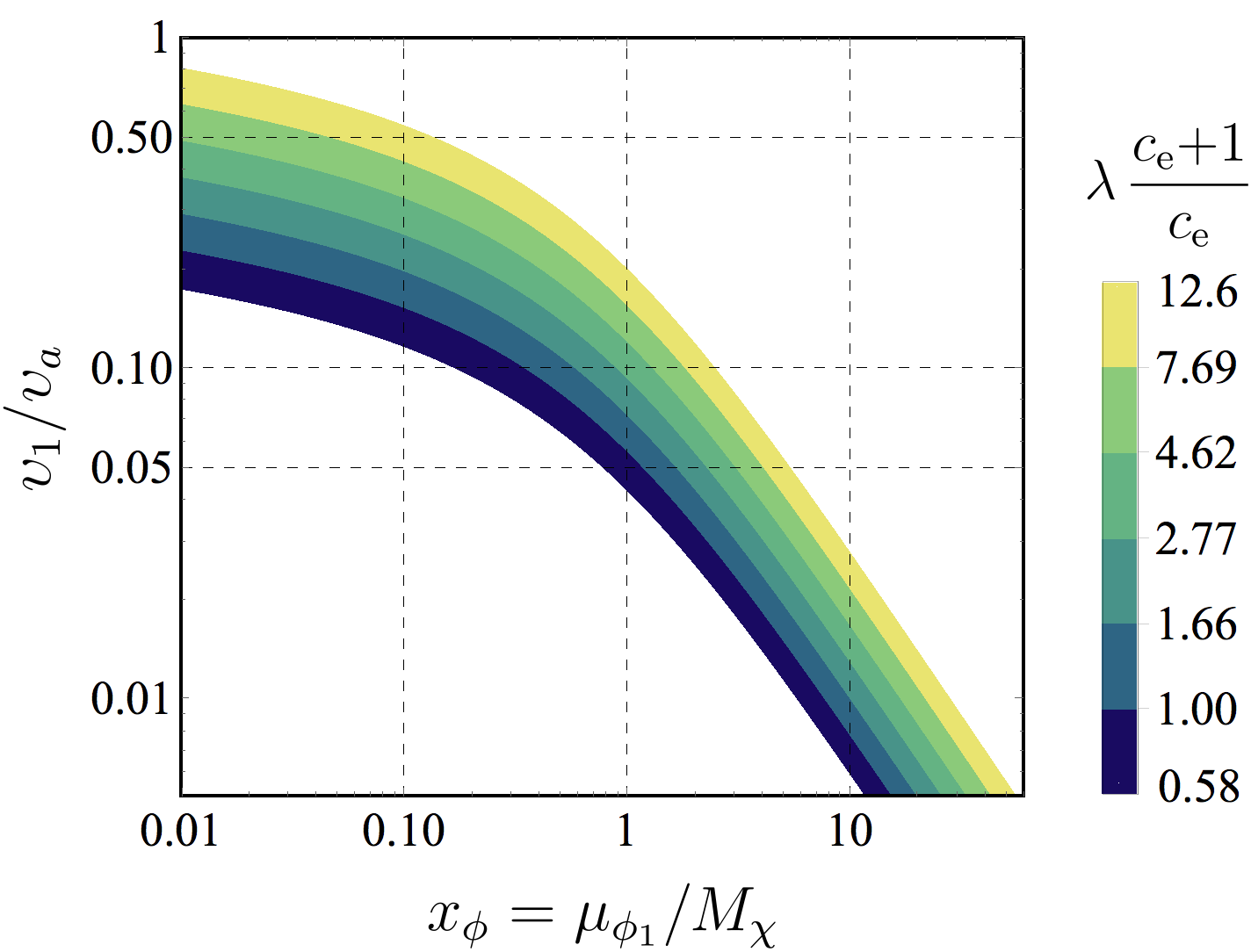}
        \caption{\label{fig:v1v2xphi} Ratio $\upsilon_1/\upsilon_a$ as function of $x_\phi$ for the muon case (Left panel) and the electron case (Right panel). We have used Eq.(\ref{eq:mRAD_vs_mFN}) with $1/2<m_\mu^{\rm FN}/m_\mu^{\rm RAD}<9/10$ and $10/9<m_\e^{\rm FN}/m_\e^{\rm RAD}<2$ and $\lambda\in [\pi/6,2\pi]$. For a given $c_\ell$, we can read the required value of $\lambda$ from the bar legend.}
\end{figure}

In the mechanism described so far, the realization of the $(g-2)$ correction is directly related to the mass generation through a flavor symmetry. This implies that ${\rm Sign}(\Delta a_\ell)={\rm Sign}(m_\ell^{\rm RAD})$ as they come from the same diagram.
In this framework, the obstacle of obtaining the experimental sign difference, between $\Delta a_\mu$ and $\Delta a_\e$, can be nicely overcome.
We can achieve it by requiring ${\rm Sign}(\Delta a_\mu)={\rm Sign}(m_\mu)$ while ${\rm Sign}(\Delta a_\e)=-\,{\rm Sign}(m_\e)$. The rotation to the physical basis where $m_\e,m_\mu >0$ automatically gives $\Delta a_\e<0$ and $\Delta a_\mu>0$. {In particular, the SM contribution to $a_e$ will get the chirality change through the electron mass itself, including both tree and radiative contributions to the mass, while our new contribution gets the chirality change through the radiative contribution only, with negative sign after rephasing.}
As the electron and muon masses are generated through a destructive interference between the FN and the radiative processes in Figure~\ref{fig:Masses}, basically what we need is an opposite cancellation in the muon and electron sectors, i.e. $\murad >\muFN $ while $\merad<\meFN $. Now, taking $g_\ell$, $\upsilon_1$ and $\varepsilon$ positive in Eq.(\ref{eq:MassFN}), this means that the radiative masses and anomalous moments, as the muon mass before rephasing, are negative while the electron mass is positive. Altogether, this implies that the radiative and the tree level contributions must be of the same order.
We let $m_\ell^{\rm RAD}$, and consequently $m_\ell^{\rm FN}$, to be up to one order of magnitude larger than $m_\ell^{\rm exp}$, i.e.
\bea \label{eq:FT}
| m^{\rm RAD}_\ell | &=&c_{\ell}\,m^{\rm exp}_\ell \hspace{5mm}{\rm with }\hspace{5mm} c_\ell\:\in\: [\,1 , 10\,]\,.
\eea
Notice that in the following ratios the dependence on the variables of our mechanism is to great extent simplified \cite{Okada:2013iba},
\bea 
\label{eq:Da_vs_mRAD}
\frac{\Delta a_\ell}{m_\ell^{\rm RAD}}&=& \frac{|\Delta a_\ell|}{c_\ell\,m_\ell^{\rm exp}}=\frac{2\, {m^{\rm exp}_\ell}}{M_\chi^2}\,\frac{I_{\Delta a}^\times(x_\phi)}{I_{m}^\times(x_\phi)}\,,\\
\label{eq:mRAD_vs_mFN}
\left| \frac{\lambda\,m_\ell^{\rm FN}}{m_\ell^{\rm RAD}} \right|&=& \lambda \frac{c_\ell \pm 1}{c_\ell} =\frac{ 16 \pi^2}{I_{m}^\times(x_\phi)}\,\left(\frac{\upsilon_1}{\upsilon_a}\right)^2,
\eea~with $(+)$ for the electron and $(-)$ for the muon, where, as before, we take $\lambda >0$. As we announced before, the ratios $m^{\rm FN}/m^{\rm RAD}$ and  $\Delta a/m^{\rm RAD}$ do not depend on the choice of $g_\ell$, as the same couplings necessarily enter the three observables. From Eq.(\ref{eq:Da_vs_mRAD}) we can directly deduce the dependence of $M_\chi(x_\phi)$ once we impose the experimental bounds on $\Delta a_\ell^{\rm exp}$ together with $c_\ell=[1,10]$, as shown in Figure~\ref{fig:MXxphi}. On the other hand, in Figure~\ref{fig:v1v2xphi}, we can see that Eq.(\ref{eq:mRAD_vs_mFN}) gives the relation of $\upsilon_1/\upsilon_a(x_\phi)$. From Figure~\ref{fig:v1v2xphi} we see that it is always true that $\upsilon_a > \upsilon_1$, as expected from the previous discussion. Notice that, although a quartic coupling $(\phi_1^\dagger \phi_1)^2$, present in Eq.(\ref{eq:Vphi}), also closes the loop in Figure~\ref{fig:Masses}, these results demonstrate the need of the non-trivial quartic coupling $(\phi_a^\dagger \phi_1)^2$ in the scalar potential. The results shown in Figure~\ref{fig:MXxphi} and \ref{fig:v1v2xphi} rely exclusively on the level of cancellation between the FN and radiative diagrams, therefore they can be considered to some extent independent of the model details. Nonetheless, their validity can only be established within a specific flavor model. For instance, the required relation between $\mu_{\phi_1}$ and $\upsilon_{(1,a)}$ will be allowed only for certain regions of the viable parameter space. Our results show superposition over the muon and electron parameter space for $M_\chi\in[0.6,2.5]$ TeV. Consequently, we use the mechanism described in this section to build a toy model based on a $U(1)_f$ flavor symmetry that accommodates both the muon and electron $(g-2)$ anomalies.

\section{A \texorpdfstring{\boldmath $U(1)_f$}{Lg} toy model}
\label{sec:toy}
 To give an illustrative realization of the mechanism described in the previous section, let us consider an Abelian flavor symmetry $U(1)_f$ generating the flavor structures. The field charges, supplemented by the appropriate mediator sector, are specified in Table \ref{tab:charges}. Here we do not consider the flavor structures involving the $\tau$, as it goes beyond our exemplifying purposes. 
\begin{table}[t!]
    \centering
    \begin{tabular}{|c c c c c c c c c c c|}
    \hline
    \bf Field &  \boldmath $\mu_L$ & \boldmath$\mu_R$ & \boldmath$\e_L$ & \boldmath$\e_R$ & \boldmath$\chi_R$ & \boldmath$\phi_1$ & \boldmath$\phi_3$ & \boldmath$\phi_a$ & \boldmath$\phi_b$ & \boldmath$H$ \\[3pt]
    \hline
       $U(1)_f$ & $-2$ & $0$ & $8$ & $3$& $1,2 \dots 6,7,8$ & $1$ & $3$ & $2/5$ & $8/5$ & 0\\[3pt]
       $Z_2$ & $+$ & $+$ & $-$ & $+$& $\pm$ & $-$ & $-$ & $+$ & $+$ & +\\[3pt]
    \hline
    \end{tabular}
    \caption{ \label{tab:charges} Fields and their flavor symmetry assignments.}
\end{table}
Apart from flavor charges, all flavons are SM singlets and mediators have the quantum numbers of lepton singlets, while the SM Higgs boson does not transform under the flavor symmetry.

We do not contemplate the presence of mediators of fractional charge. This is a crucial assumption, as it forbids the possibility for $\phi_{(a,b)}$ to participate to the mass generation at tree-level through the FN mechanism. Besides, two distinct fields $\phi_a$ and $\phi_b$ are required if they must have fractional charges. A term $(\phi_a^\dagger \phi_1)^2$, as in Eq.~(\ref{eq:Vphi}), would require $(2 q_1 - 2 q_a =0)$ and hence the same charge as $\phi_1$. 
Furthermore, in this model we introduce other two different flavons, $\phi_1$ and $\phi_3$, to obtain different cancellations between $m^{\rm FN}$ and $m^{\rm RAD}$ for the electron and the muon. If we have a single flavon, $\phi_1$, it is easy to see that $m^{\rm FN}/ m^{\rm RAD}$ is the same for both electrons and muons. Moreover, we need this ratio to be negative to obtain a cancellation. As we will see below, both conditions are met with the introduction of $\phi_3$.

The $Z_2$ symmetry plays a fundamental role. Any diagram that couples $\ell^{+}\rightarrow\ell^{+(-)}$, where superscripts refer to $Z_2$ charges, requires an even (odd) number of insertions. As we consider only flavons with odd charges, our choice of $U(1)$ charges could allow $\e_R^+\rightarrow\mu_L^+$ only at the level of $(2n+1)$-insertions and $\mu_R^+\rightarrow\e_L^-$ with $(2n)$-insertions. However the $Z_2$ symmetry prevents any of these flavor-changing couplings that would give rise to $\mu\rightarrow \e \gamma$. For the same reason, it  also eliminates any effective vertex $\mu_R^\dagger\e_R$ and $\mu_L^\dagger\e_L$. Thus the charge assignments in Table~\ref{tab:charges} conserves leptonic flavors. 

The effective Lagrangian preserving the charge assignment of the underlying $U(1)_f$ flavor symmetry has the form
\bea
\label{eq:Lagrangian}
\mathcal{L}_\ell&=&
g_\mu \left[\mu_R^{(0)}\overline{\chi}_R^{(1)}\phi_1^\dagger +\overline{\mu}_L^{(2)}\chi_R^{(-2)} H^{(0)}\right]+
g_\e\left[ \e_R^{(3)}\overline{\chi}_R^{(-6)}\phi_3+\overline{\e}_L^{(8)}\chi_R^{(-8)}H^{(0)}\right]\\
&+& \,g\sum_q \left[ \overline{\chi}^{(q)}_R\chi_R^{(q+1)}\phi_1+ \overline{\chi}^{(q)}_R\chi_R^{(q+3)}\phi_3 \right]  \rm\,+\, h.c.\quad.
\eea
\begin{figure}[t!]
  \centering
  \includegraphics[scale=0.58]{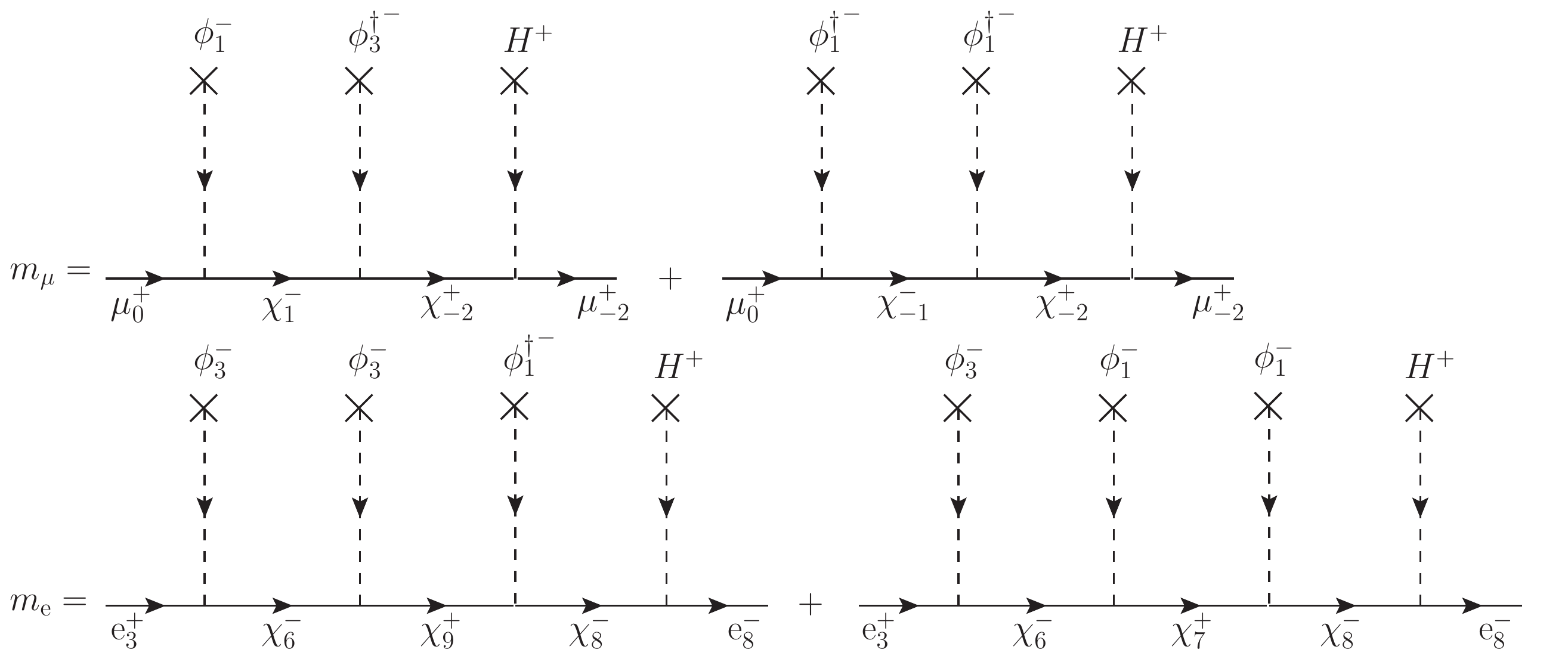}
  \caption{\label{fig:mFNtoy} FN diagrams entering in the generation of the electron and muon masses. The field subscripts indicate the $U(1)_f$ charge while the superscripts specify the $Z_2$ assignments.}
\end{figure}
Given the charge assignment in Table \ref{tab:charges}, the most general scalar potential can be written as
\bea \label{eq:toyV}
    V & = & \mu_1^2\, \phi_1^{^\dagger} \phi_1 \;+\; \mu_3^2\, 
	    \phi_3^\dagger \phi_3 \;+\; \mu_a^2\, \phi_a^{\dagger} \phi_a \;+\; 
        \mu_b^2\, \phi_b^{\dagger} \phi_b \nn \\
		& + & 	\lambda_{1}\, \left(\phi_1^\dagger \phi_1 \right)^2 \;+\;
					\lambda_{3}\, \left(\phi_3^\dagger \phi_3 \right)^2 \;+\; 
					\lambda_{a}\, \left(\phi_a^{\dagger} \phi_a \right)^2 \;+\; 
					\lambda_{b}\, \left(\phi_b^{\dagger} \phi_b \right)^2 \nn \\	
		& + & 	\lambda_{13}\, \left(\phi_1^{\dagger} \phi_1 \right) \left( \phi_3^\dagger \phi_3\right) \;+\;
					\lambda_{1 a}\, \left(\phi_1^{\dagger} \phi_1 \right) \left( \phi_a^{\dagger} \phi_a\right) \;+\; 
					\lambda_{1b}\, \left(\phi_1^{\dagger} \phi_1 \right) \left( \phi_b^{\dagger} \phi_b\right) \\
		& + & 	\lambda_{3 a}\, \left( \phi_3^\dagger \phi_3\right) \left(\phi_a^{\dagger} \phi_a \right) \;+\;
					\lambda_{3 b}\, \left( \phi_3^\dagger \phi_3\right) \left(\phi_b^{\dagger} \phi_b \right)\;+\;
					\lambda_{ab}\, \left( \phi_a^{\dagger} \phi_a\right) \left(\phi_b^{\dagger} \phi_b \right) \nn \\
		& + & 	\left(\lambda_{1ab}\,\phi_a^\dagger \phi_b^\dagger \phi_1^2+ \lambda'_{13}\,\phi_3^\dagger \phi_1^3+ {\mu'_a}^{2}\phi_a^2+ {\mu'_b}^{2}\phi_b^2\nonumber +{\rm h.c.}\right), 
\eea
where the last two terms are introduced to break explicitly the $U(1)_f$ symmetry and allow to give a small mass to the two Goldstone bosons\,\footnote{It is easy to check that taking ${\mu}^\prime_a ={\mu}^\prime_b = 0$, the potential has two unconstrained charges and therefore two global symmetries. Initially, we have four charges $q_1$, $q_3$, $q_a$ and $q_b$, {\it i.e.} 4 symmetries. Then, only the last row in Eq.~(\ref{eq:toyV}) constrains these charges, $(2 q_1-q_a-q_b = 0)$ and $(3 q_1-q_3 = 0)$. So, there remain two global symmetries that are explicitly broken by $\mu'_a$ and $\mu'_b$. As can be seen in the Appendix, we have two pseudoscalar masses directly proportional to $\mu_{(a,b)}^{\prime\,2}$.}
present in the model, while the $Z_2$ is preserved.\footnote{{As mentioned in Section \ref{sec:theory}, neutrino masses can be accommodated through the right-handed neutrino Majorana masses. The breaking of $Z_2$ would be produced by the same flavons breaking lepton number, coupling only to $\nu_R$. This allows $\mu-e$ mixing in the $\nu_R$ and, hence, in the $\nu_L$ mass matrices. In this way, the charged-lepton sector would be practically unaffected, with flavor changes in charged-leptons always proportional to  neutrino masses.}} For simplicity we consider the $\lambda$s to be real. The flavor symmetry is spontaneously broken when the flavons get a nonzero vev at the minimum of the scalar potential. As detailed in Appendix \ref{AppSec:Vmin}, the potential in Eq.(\ref{eq:toyV}) allows for a non trivial minimum with $\upsilon_3\sim -2\upsilon_1$ and $\upsilon_b \sim \upsilon_a$, $\upsilon_{(1,a)}\neq 0 $ and $\upsilon_1<\upsilon_a$.
The mass matrices of the CP-even ($S_i$) and -odd bosons ($P_i$) can be diagonalized by two orthogonal matrices as detailed in Appendix \ref{AppSec:Vmin}. The relevant (pseudo) scalar masses are 
\bea
\label{eq:M2eig}
m_{S_1}^2 &\simeq& 2\upsilon_1^2\left(2\lambda_1 - \lambda_{13}-\cfrac{9}{4}\lambda'_{13}\right) \hspace{6mm},\hspace{2mm}
m_{P_1}^2 \simeq -2\,\upsilon_a^2\left(\lambda_{1ab}+\frac{\upsilon_1^2}{2\upsilon_a^2}(\lambda_{1ab}-18\lambda'_{13})\right)\,,\\
m_{S_2}^2 &\simeq& 2\,\upsilon_1^2\left(2\lambda_1+4\lambda_{13}-6\lambda'_{13} - \cfrac{5}{4} \cfrac{(2\lambda_{1a}+\lambda_{1ab})^2}{2\lambda_{a}+\lambda_{ab}}\right)\,.
\eea
These physical masses are related to the $\mu^2_{\phi_1} \equiv \mu^2_1$ in Figure~\ref{fig:MXxphi} as $m^2_{S_{1,2}}\sim (2\lambda_{1a}+\lambda_{1ab})\upsilon^2_a + 6 \lambda_1 \upsilon_1^2  - \mu_{\phi_1}^2$ and   $m^2_{P_{1}}\sim (2\lambda_{1a}-\lambda_{1ab})\upsilon^2_a + 6 \lambda_1 \upsilon_1^2 - \mu_{\phi_1}^2$, relations that are valid up to $\mathcal{O}(\upsilon^2_1/\upsilon^2_a)$ corrections. 
Looking at $m^2_{P_1}$, it is clear that the necessary condition for a minimum is $\lambda_{1ab}<0$. 

\subsection[Mass generation and $(g-2)_\ell$]{Mass generation and \texorpdfstring{\boldmath $(g-2)_\ell$}{Lg}}
\label{sec:ToyMass}
Using the vertices in Eq.(\ref{eq:Lagrangian}), we can write down the FN diagrams entering in the mass generation of $m_\mu$ and $m_\e$. They are displayed in Figure~\ref{fig:mFNtoy}, where it is important to notice that due to the presence of $\phi_3$ we have different tree-level diagrams contributing to $m_\mu$ and $m_e$ with different weights for  $\upsilon_3\sim -2\upsilon_1$.  From the potential in Eq.(\ref{eq:toyV}) we see that different quartic couplings can act closing the loop in one of these diagrams for $m_\mu$ and $m_e$.  In our toy model, the $\lambda$ introduced in Eqs.(\ref{eq:RadMass},\ref{eq:(g-2)}) is given by the sum of different terms
\bea
\lambda \rightarrow \lambda_{1ab}\frac{\upsilon_{b}}{\upsilon_{a}}+\lambda_{1}\frac{\upsilon_{1}^2}{\upsilon_{a}^2} +\lambda'_{13}\frac{\upsilon_3\upsilon_1}{\upsilon_a^2}.
\eea
Nevertheless, the $\phi_{(1,3)}$ couple directly to the SM fermions and the size of their vevs are limited, while the $\upsilon_{(a,b)}$ only enter the masses at loop level and their values can be correspondingly larger. Provided that $\upsilon_{a}\sim \upsilon_{b}\gg \upsilon_{(1,3)}$, only diagrams with two or more $\phi_1$, closed by the quartic coupling $\lambda_{1ab}\,\phi_a^\dagger\,\phi_b^\dagger\phi_1^2$ can give a contribution to $(g-2)_\ell$ with the required enhancement. Then, the total masses are,
\bea
\label{eq:mmutot}
m_\mu &=& g_\mu^3\frac{\upsilon_H}{\sqrt{2}}\varepsilon_1^2\left[\left(\frac{\varepsilon_3}{\varepsilon_1}+1\right) + \frac{\lambda_{1ab}}{16 \pi^2}\frac{\varepsilon_a^2}{\varepsilon_1^2} I_{m}^\times(x_\phi)\right]\sim g_\mu^3\frac{\upsilon_H}{\sqrt{2}}\varepsilon_1^2\left[-1 + \frac{\lambda_{1ab}}{16 \pi^2}\frac{\varepsilon_a^2}{\varepsilon_1^2} I_{m}^\times(x_\phi) \right]\!,~~\\
\label{eq:metot}
m_\e &=& g_\e^4\frac{\upsilon_H}{\sqrt{2}}\varepsilon_1^2 \varepsilon_3\left[\left( 2\,\frac{\varepsilon_3}{\varepsilon_1}+1\right) + \frac{\lambda_{1ab}}{16 \pi^2}\frac{\varepsilon_a^2}{\varepsilon_1^2} I_{m}^\times(x_\phi)  \right]\sim 2 \,g_\e^4\frac{\upsilon_H}{\sqrt{2}}\varepsilon_1^3 \left[  3 -\frac{\lambda_{1ab}}{16 \pi^2}\frac{\varepsilon_a^2}{\varepsilon_1^2} I_{m}^\times(x_\phi)  \right]\!,~~
\eea
 where $\varepsilon_{(1,a)}=\upsilon_{(1,a)}/\mX$, we assume a common mediator mass $M_\chi$ to simplify the discussion and, in the second equality, we have taken $\upsilon_3\sim -2 \upsilon_1$. In this equation we can see that, as we said above, it is the presence of $\upsilon_3$ which provides the negative relative sign and different cancellation in $m_\mu$ and $m_e$.  Now, the corresponding contributions to $(g-2)_\ell$ read as
\bea 
\label{eq:toy(g-2)mu}
\Delta a_\mu & \sim & g_\mu^{3}\,\frac{\lambda_{1ab}}{8\, \pi^2}\,
                        \frac{\upsilon_H}{\sqrt{2}}\,\frac{m^{\rm exp}_\mu}{\mX^2}\,
                        \varepsilon_a^2\,I_{\Delta a}^{\times}(x_\phi)\,,\\
\label{eq:toy(g-2)e}
\Delta a_\e & \sim & -2\,g_\e^{4}\,\frac{\lambda_{1ab}}{8\, \pi^2}\, 
                        \frac{\upsilon_H}{\sqrt{2}}\,\frac{m^{\rm exp}_\e}{\mX^2}
                        \varepsilon_a^2 \,\varepsilon_1\,
                        I_{\Delta a}^{\times}(x_\phi)\,.
\eea
The minimization of the scalar potential requires $\lambda_{1ab}<0$ and the loop function is also $I_m^\times(x_\phi)<0$, so the radiative diagram gives a positive contribution to the mass. 
From Eqs.(\ref{eq:toy(g-2)mu}) and (\ref{eq:toy(g-2)e}) one sees that, to obtain ${\rm Sign}(\Delta a_\mu)=-{\rm Sign}(\Delta a_\e)$ in the physical basis, we need the following condition to be satisfied 
\bea
\frac{1}{\sqrt{3}} <\frac{4\pi}{\sqrt{\lambda_{1ab} I_m^\times (x_\phi)}} \frac{\varepsilon_1}{\varepsilon_a} <1\,.
\eea

\begin{table}[t!]
{\small
    \centering
    \setlength{\tabcolsep}{4pt}
    \begin{tabular}{|c c c c c c c c c | c c c |c|}
    \hline
    \boldmath$M_\chi$   &  \boldmath$m_{S_1}$ & \boldmath$m_{S_2}$ & \boldmath$m_{S_3}$ & \boldmath$m_{S_4}$ & \boldmath$m_{P_1}$ & \boldmath$m_{P_2}$ & \boldmath$m_{P_3}$ & \boldmath$m_{P_4}$ & \boldmath$\upsilon_1$ & \boldmath$\upsilon_3$ & \boldmath$\upsilon_{(a,b)}$ &  \boldmath$g_\e$  \\[3pt]
    1658 & 123 & 337 & 1245 & 1430 & 611 & 23 & 18 & 18 & 42 & -84 & 262 &  0.72\\[3pt]
    \hline
     \boldmath$\lambda_{1}$ & \boldmath$\lambda_3$ & \boldmath$\lambda_{(a,b)}$ & \boldmath$\lambda_{13}$ & \boldmath$\lambda_{1(a,b)}$ & \boldmath$\lambda_{3(a,b)}$ & \boldmath$\lambda_{ab}$ & \boldmath$\lambda'_{13}$ & \boldmath$\lambda_{1ab}$ &\boldmath$\mu_{(1,3)}$ & \boldmath$\mu_{(a,b)}$ & \boldmath$\mu'_{(a,b)}$  & \boldmath$g_\mu$  \\[3pt]
    5.93 & 3.31 & 6.54 & 6.08 & 0.97 & -0.31 & 1.82 & 0.65 & -2.50 & 122 & 1010 & 9 & 0.85  \\[3pt]
    \hline
    \end{tabular}
    \caption{ \label{tab:benchmark} Example of a benchmark point. The spectrum mass parameters are given in GeV. The combination of parameters provides $\Delta a_\mu = 1.6 \times 10^{-9}$ and $\Delta a_\e = -1.8 \times 10^{-13}$ with a relative size of the loop contributions $(c_\e,c_\mu)=(7.1,1.6)$. }
}
\end{table}
 An example of a set of numerical values of the parameters
 giving a global minimum, the corresponding vevs, and the
 resulting scalar mass spectrum are shown in Table \ref{tab:benchmark}. Notice that as expected there are two light pseudoscalars, i.e.~the pseudo Nambu-Goldstone bosons, with mass of the order of the explicit $U(1)_f$ breaking,
 and a third pseudoscalar which is instead light because its
 mass is controlled by the small vev $\upsilon_1$.
 
\section{Phenomenological implications}
\label{sec:Pheno}
We have seen that to explain the discrepancies in the muon and electron anomalous magnetic moments through a low scale flavor symmetry, a relatively light flavon and mediator sector is required. 
In this section we discuss the phenomenology of these light particles at colliders and precision experiments. Rather than focusing on the specific toy model presented in \sref{sec:toy}, we discuss the general features and phenomenological consequences of the mechanism outlined in \sref{sec:theory}.

In Figures~\ref{fig:MXxphi} and \ref{fig:v1v2xphi}, we can see the requirements on the masses and the vacuum expectation values needed to reproduce the anomalous magnetic moments through this mechanism, irrespective of the details of the model, as symmetries, charges, and scalar potential. The figure shows that we can successfully reproduce $(g-2)_\mu$ at the 2$\sigma$ level with a mediator mass up to 5.7 TeV, although this implies that $c_\mu = 10$, i.e.~a cancellation of the tree-level and radiative contributions to the muon mass with a tuning of 10\%. In the case of $(g-2)_\e$ at 2$\sigma$ the maximum allowed mediator mass is 2.5 TeV with a 10\% tuning. 

If we take both values at 2$\sigma$, we can see that we relax both the electron and muon discrepancies with $M_\chi \simeq 2.5$~TeV and $x_\phi \simeq 0.6$. This implies $c_\e = 10$ and $c_\mu  \in [2.2,6.9]$, where the $c_\mu$ range reflects the 2$\sigma$ range in \eref{eq:dam}. Then, the cancellation is larger for the electron that for the muon and, as expected, a smaller degree of cancellation would imply a lighter mediator. For instance, to reproduce the central values with $M_\chi = 1$~TeV and $x_\phi = 1$, it would require $c_\mu = 1.1$ and $ c_e = 7.7$. Therefore, our explanation of the muon and electron discrepancies in the anomalous magnetic moments at two sigmas has a definite prediction: we expect vector-like fermions with the quantum numbers of right-handed and/or left-handed SM leptons with mass below 2.5~TeV. 

The scalar sector is more model dependent, as the exact spectrum depends on the minimization of the scalar potential as exemplified in \aref{AppSec:Vmin} for the toy model. We can however outline some general features, based on the discussion in \sref{sec:theory}. Figures~\ref{fig:MXxphi} and \ref{fig:v1v2xphi} show that for our mechanism to work we need: (i) a hierarchy between the $U(1)_f$-breaking vevs with those  (``$\upsilon_1$'') entering the tree-level mass diagrams smaller than those (``$\upsilon_a$'') controlling the radiative mass and the contributions to $(g-2)_\ell$, i.e.~$\upsilon_1 < \upsilon_a$; (ii) the bilinear terms in the scalar potential $\mu_\phi$ of the flavons coupling to leptons of the same order or smaller than the mediator mass $M_\chi$, unless $\upsilon_1 \ll \upsilon_a$.
It is thus reasonable to expect at least one scalar and/or pseudoscalar to be much lighter than the mediators. 
This is indeed the case in the explicit example shown in Table \ref{tab:benchmark}, where the scalar spectrum lies in the 10\,GeV--2\,TeV range. The light states have in particular to come mostly from the flavons involved in the FN diagram, thus coupling to light leptons, that in \sref{sec:theory} we denoted as $\phi_1$.
Besides, there must be one or more pseudo-Goldstone bosons whose mass is controlled by explicit $U(1)_f$-breaking terms
and thus naturally\,---\,although not necessarily\,---\,light.

Given the above discussion, here we focus on the phenomenology of scalar states with a substantial component of the flavon $\phi_1$ entering the tree-level FN diagrams that are in general expected to have mass of $\mathcal{O}(100)$ GeV or lighter.

From \fref{fig:Masses}, one can see that the coupling $y_{\phi\ell}$ of a physical state in $\phi_1$ to $\ell_L\ell_R$ 
is proportional to the FN contribution to the lepton mass:
\begin{align}
y_{\phi\ell} \approx n_\phi \frac{m^{\rm FN}_\ell}{\upsilon_1}, 
\end{align}
where $n_\phi$ is the number of $\phi_1$ insertions in the diagram. 
Even considering the maximal tuning we allowed, $m^{\rm FN}_\ell = 10\, m_\ell^{\rm exp}$ (i.e.~$c_\ell=10$), the ratio 
${m^{\rm FN}_\ell}/{\upsilon_1}$ provides a substantial suppression to the couplings to electrons and muons.
Indeed, numerically the couplings result
\begin{align}
\label{eq:yphiel}
y_{\phi \e} ~\approx ~ 2\times10^{-4} \, \left(\frac{n_\phi}{2}\right)\left(\frac{c_\e}{10}\right)   \left(\frac{50\,\rm GeV}{\upsilon_1}\right) , \quad  \\
y_{\phi \mu} ~\approx ~4\times10^{-2} \, \left(\frac{n_\phi}{2}\right)\left(\frac{c_\mu}{10}\right)   \left(\frac{50\,\rm GeV}{\upsilon_1}\right).
\end{align}
The flavon couples preferably to the heaviest lepton, in our case the muon. Of course, it would be the tau if the same flavon were involved in the generation of the tau mass. As a consequence, if produced at colliders either directly or through decays of the mediators, our flavon would decay as $\phi_1 \to \mu^+\mu^-$ (or $\tau^+\tau^-$) with a branching ratio close to 100\%. 
A $\phi_1$ lighter than about 200 GeV could appear as a di-muon (or di-tau) resonance at LEP: 
$\e^+ \e^- \to \phi_1\to \mu^+\mu^-$. However, the production cross section depends on the small coupling to electrons and, due to limited statistics, searches for such kind of di-fermion resonances performed by LEP experiments are not sensitive to
couplings $y_{\e\phi} \lesssim 10^{-2}$ \cite{Tanabashi:2018oca}. For flavons substantially heavier than the maximum LEP center-of-mass energy (209 GeV), bounds on the 4-lepton contact interaction \cite{Schael:2013ita} translate into a limit $y_{\e\phi}\,  y_{\mu\phi} \lesssim 5\times 10^{-3}~ (m_{\phi}/400\,\rm GeV)^2$, several orders of magnitude above our typical values shown in \eref{eq:yphiel}.
It would be interesting to assess the sensitivity of proposed future leptonic colliders\,---\,such as the ILC, CLIC, CEPC, and FCC-ee, see e.g.~\cite{Strategy:2019vxc}\,---\,to leptonic flavons, a question that we defer to future work.

The FN mediators we considered are heavy vector-like leptons with the quantum numbers of the SM lepton singlets, although 
realizations of our mechanism involving also or exclusively SU(2) doublet mediators are conceivable.  
In either case, these new heavy fermions can be abundantly produced at the LHC via the electro-weak Drell-Yan process 
$pp\to Z^{*}/\gamma^* \to \chi^+ \chi^-$, plus modes involving the neutral states in case of doublet mediators.
In general, vector-like leptons mix with the SM leptons, hence the charged states can decay to light leptons and SM bosons: 
$\chi^\pm \to Z\,(h)\,\ell^\pm$,  see e.g.~\cite{Kumar:2015tna}. In our case a more direct decay mode involves lighter flavon states: $\chi^\pm \to \phi_1 \,\ell^\pm$, where again with $\phi_1$ we denote a flavon appearing in FN diagrams. Depending on the FN charge of a given mediator, decays of this kind may occur through a renormalizable $\mathcal{O}(1)$ coupling, or again through an effective coupling arising from mixing of different mediators involving the insertion of a certain number of flavon and/or Higgs, as one can see from FN diagrams such as in \fref{fig:mFNtoy}. As in general a fewer number of vev insertion is needed than for the decays to SM particles, we expect that this mode will be always dominant. The exact quantum numbers of a given mediator will also determine which lepton the mediator preferably decays into. Considering that as discussed above flavons decay to pairs of the heaviest lepton they couple to, the typical signature of this kind of models at the LHC consists of a multi-lepton final state such as:
\begin{equation}
pp\to \chi^+ \chi^- \to \phi_1(\to \mu^+\mu^-)\, \ell^+\, \phi_1(\to \mu^+\mu^-)\, \ell^-,
\label{eq:VLprod}
\end{equation}
where  $\ell = \e,\,\mu$ and the di-muon invariant mass can reconstruct the mass of the $\phi_1$ state. Of course for models involving the third generation, decay chains of this kind involving taus are possible and, in particular, flavons coupling to a FN diagram for the tau would mostly decay into $\tau^+\tau^-$.

Searches based on multi-lepton final states have been performed by the LHC collaborations \cite{Aad:2015dha,Aaboud:2018zeb,Sirunyan:2019ofn,Sirunyan:2019bgz}, and employed to constrain a variety of new-physics models. In particular the analysis in  \cite{Sirunyan:2019ofn} was interpreted in terms of production of third generation vector-like lepton doublets
decaying to SM gauge/Higgs bosons and taus/tau neutrinos. A limit on the mass of the vector-like lepton $\gtrsim 800$ GeV was obtained. We expect that reinterpreting this and other multi-lepton searches in terms of the vector-like lepton production and decay chain shown in \eref{eq:VLprod} would yield a comparable limit, possibly stronger, in the 1 TeV ballpark, if no taus or neutrinos are present in the final state. An optimized search taking full advantage of the spectacular six-lepton signature in \eref{eq:VLprod} should further increase the sensitivity. 

Finally, we conclude this section by commenting about possible low-energy probes of our setup.
The most obvious observables that could test a combined explanation of both electron and muon $g-2$ are LFV processes and the electron EDM. In fact, the suppression of LFV processes does not need to be complete as in the toy model of \sref{sec:toy},
and any deviation from a perfect flavour alignment of the dipole coefficients $C_{\ell\ell^\prime}$ in \eref{eq:L-dipole} could be observed by searches for LFV processes, cf.~\cite{Calibbi:2017uvl} for status and prospects of these experiments. 
The same diagrams giving rise to $(g-2)_\e$ can contribute to the electron EDM (eEDM). In terms of the usual effective operators such contribution reads
\begin{align}
d_{\e} =  \frac{e \,m_\e}{4\pi^2}  \,{\rm Im}(C_{\e\e}).
\end{align}
The latest experimental limit \cite{Andreev:2018ayy} then implies:
\begin{equation}
d_\e < 1.1\times 10^{-29}~e\,{\rm cm}~\Rightarrow~|{\rm Im}(C_{\e\e})| < 6 \times 10^{-7} ~{\rm GeV}^{-2}.
\end{equation}
Comparing this with \eref{eq:Cee}, we can see that the suppression of the imaginary part of $C_{\e\e}$, thus of the overall CP-violating phase of the $(g-2)_\e$ diagram, with respect to the real part must be at the percent level. 
Therefore, unless the CP-violating phase is exactly zero, as it is the case if all new couplings are real, the eEDM is an observable where a non-standard $(g-2)_\e$ can be tested, cf.~a related discussion in \cite{Crivellin:2018qmi}.

\section{Conclusions}
\label{sec:Conclusions}
We have proposed a new mechanism to accommodate the experimental $(g-2)_{\ell}$ ($\ell = \e,\mu$) discrepancies within the framework of low-scale flavor symmetry models. In these flavour models, that generate the Yukawa couplings through a Froggatt-Nielsen mechanism, the presence of quartic couplings between flavons can always act to close the loop of two scalar flavons that contribute to the mass at tree level, and thus both give a radiative correction to the mass and generate a contribution to the magnetic moment. We stress that a sizeable contribution of the anomalous magnetic moment, as required by the observed discrepancies, gives necessarily a contribution to the mass. 

In order to obtain a sizable $g-2$ correction, compatible with the present discrepancies, we introduce a nontrivial quartic coupling with a second flavon, that acquires a large VEV though does not participate to the tree level masses. The radiative mass receives the same enhancement and contributes significantly to the mass generation; this sets a limit on the size of the $g-2$ contribution. The FN and radiative diagrams, with opposite signs, contribute to the electron and muon masses through a cancellation that accommodates the experimental difference in sign between the electron and muon magnetic moment discrepancies and, at the same time, contributes to satisfy the experimental limit on searches of vector-like mediators. 

We show that our mechanism can provide a simple explanation of the discrepancies of the muon $(g-2)_\mu$ and the electron $(g-2)_\e$, simultaneously in a large viable parameter space, with predicted mediator masses as large as $M_\chi \in [0.6, 2.5]$ TeV. We give an example of how this can be achieved in a toy model based on a $U(1)_f$ flavor symmetry. The application to a complete model, including the tau, quarks and neutrino sectors and the study of its phenomenological consequences in flavor physics is left to future works.

\appendix
\section{Minimization of the potential}
\label{AppSec:Vmin}
In order to reduce the number of free parameters, we consider the following relations among coefficients: $\lambda_b\sim\lambda_a$, $\lambda_{(1,3)b}\sim\lambda_{(1,3)a}$, $\lambda_{3 a} \sim (\lambda_{1a}+\lambda_{1ab})$, $ \lambda_3 \sim (4\lambda_1 + 6 \lambda_{1 3} - 11 \lambda'_{1 3})/16$, $\mu_1\sim\mu_3$, $\mu_b\sim\mu_a$, $\mu'_b\sim\mu'_a$. They are a total of 8 relations that reduce to 10 the number of free parameters in Eq.(\ref{eq:toyV}). We choose the following representation for the scalar fields after spontaneous symmetry breaking:
\bea
\phi_i=\upsilon_i + \sigma_i + i \varphi_i.
\eea
The minimization conditions of $V$ in Eq.(\ref{eq:toyV}), with respect to $\sigma_i$ read as
{\small
\bea
\label{eq:minimization}
\left\langle \frac{\partial V}{\partial \sigma_a} \right\rangle &=& 2 \upsilon_a \left[\upsilon_a^2 \left(2\lambda_a +\frac{\upsilon_b^2}{\upsilon_a^2}\lambda_{ab}\right) + \upsilon_1^2 \left(  \lambda_{1a} + \frac{\upsilon_b}{\upsilon_a} \lambda_{1ab} \right)+  \upsilon_3^2 (\lambda_{1a}+\lambda_{1ab})-\mu_a^2-2 {\mu'_a}^2\right]=0 \nn\\
\left\langle \frac{\partial V}{\partial \sigma_b} \right\rangle &=& 2 \upsilon_b\left[  \upsilon_b^2 \left(2\lambda_a +\frac{\upsilon_a^2}{\upsilon_b^2}\lambda_{ab}\right) + \upsilon_1^2 \left(  \lambda_{1a} + \frac{\upsilon_a}{\upsilon_b} \lambda_{1ab} \right)+ \upsilon_3^2 (\lambda_{1a}+\lambda_{1ab})-\mu_a^2-2{\mu'_a}^2\right]=0 \nn\\
\left\langle \frac{\partial V}{\partial \sigma_1} \right\rangle &=&2 \upsilon_1\left[ 2\upsilon_1^2 \left( \lambda_1+\frac{\upsilon_3^2}{2\upsilon_1^2} \lambda_{13} +  3\,\frac{\upsilon_3}{2\upsilon_1} \lambda'_{13}\right) + 2\upsilon_a^2\left(\lambda_{1a}+ \lambda_{1ab}\right)-\mu_1^2\right]=0 \,, \\ 
\left\langle \frac{\partial V}{\partial \sigma_3} \right\rangle &=& 2\upsilon_3\left\{ \frac{\upsilon_3^2}{2}\left[ \lambda_1 +2\left(\frac{ \upsilon_1^2}{\upsilon_3^2}+\frac{3}{4}\right)\lambda_{13} + 2\left(\frac{\upsilon_1^3}{\upsilon_3^3}-\frac{11}{8}\right)\lambda'_{13}\right]+2 \upsilon_a^2\left(\lambda_{1a}+ \lambda_{1ab}\right) -\mu_1^2 \right\}=0 \nonumber
\eea
}
where the symbol $\langle\cdots \rangle$ denotes that the fluctuating fields are taken to be zero. We obtain the required relations among vevs: $\upsilon_b \sim \upsilon_a$ and $\upsilon_3 \sim -2\,\upsilon_1$,
while $(\upsilon_1,\upsilon_a)$ in terms of the $\lambda$s are given by
{\small
\bea
\begin{pmatrix}
\label{eq:v1vamin}
\upsilon_1^2=0 & \upsilon_a^2=0 &V_0=0\\
\upsilon_1^2= \cfrac{\mu_1^2}{2\widetilde{\lambda}_1} & \upsilon_a^2=0 &V_1=-\cfrac{5}{4}\cfrac{\mu_1^4}{\widetilde{\lambda}_1}\\
\upsilon_1^2=0 & \upsilon_a^2= \cfrac{\widetilde{\mu}_a^2}{\widetilde{\lambda}_a}  & V_2=-\cfrac{\widetilde{\mu}_a^4}{\widetilde{\lambda}_a}\\
\upsilon_1^2=\cfrac{\mu_1^2-2\cfrac{\widetilde{\lambda}_{1a}}{\widetilde{\lambda}_{a}}\widetilde{\mu}_a^2}{2\widetilde{\lambda}_1-10\cfrac{\widetilde{\lambda}_{1a}^2}{\widetilde{\lambda}_{a}} } & \upsilon_a^2= \cfrac{2 \cfrac{\widetilde{\lambda}_{1}}{\widetilde{\lambda}_{a}}\widetilde{\mu}_a^2-5\cfrac{\widetilde{\lambda}_{1a}}{\widetilde{\lambda}_{a}}\mu_1^2}{2\widetilde{\lambda}_1 -10\cfrac{\widetilde{\lambda}_{1a}^2}{\widetilde{\lambda}_{a}}}  & V_3=\cfrac{-\cfrac{\widetilde{\mu}_a^4}{\widetilde{\lambda}_a}-\cfrac{5}{4}\cfrac{\mu_1^4}{\widetilde{\lambda}_1} + 5 \cfrac{\widetilde{\lambda}_{1a}}{\widetilde{\lambda}_{a} \widetilde{\lambda}_{1} } \mu_1^2\widetilde{\mu}_a^2 }{ 1-5\cfrac{\widetilde{\lambda}_{1a}^2}{\widetilde{\lambda}_{a} \widetilde{\lambda}_{1} }}
\end{pmatrix}\,,
\eea
}
where $\widetilde{\lambda}_1=\lambda_1+2\lambda_{13}-3\lambda'_{13}$, $\widetilde{\lambda}_a=2\lambda_a+\lambda_{ab}$ , $\widetilde{\lambda}_{1a}=\lambda_{1a}+\lambda_{1ab}/2$ and $\widetilde{\mu}_a^2=\mu_a^2+2{\mu'_a}^2$. The only interesting minimum for us is the non trivial case $\upsilon_1,\upsilon_a\neq 0$ with $\upsilon_1\ll \upsilon_a$, so we require $V_3$ to be a global minimum. The $4\times 4$ squared mass matrices of the CP-even and -odd bosons ($S_i$ and $P_i$) are given by  
\bea \label{eq:MhAV}
    \left({M}^2_S\right)_{ij} = \cfrac{1}{2} \frac{\partial^2 V}{\partial \sigma_i \partial \sigma_j}\hspace{5mm},\hspace{5mm}
 \left({M}^2_P\right)_{ij} = \cfrac{1}{2} \frac{\partial^2 V}{\partial \varphi_i \partial \varphi_j}\,.
\eea
Using the potential in Eq.(\ref{eq:toyV}), these matrices acquire the following form
{\small
\bea
M_S^2 \simeq
\begin{pmatrix}
2\upsilon_1^2(2\lambda_1-3\lambda'_{13}) & -\upsilon_1^2(-4\lambda_{13}+3\lambda'_{13}) & 2\upsilon_a \upsilon_1\widetilde{\lambda}_{1a} &  2\upsilon_a \upsilon_1\widetilde{\lambda}_{1a}\\
-\upsilon_1^2(-4\lambda_{13}+3\lambda'_{13}) & \cfrac{\upsilon_1^2}{2}(8\lambda_1+12\lambda_{13}-21  \lambda'_{13}) &  -4\upsilon_a \upsilon_1\widetilde{\lambda}_{1a} & -4\upsilon_a \upsilon_1\widetilde{\lambda}_{1a}\\
2\upsilon_a \upsilon_1\widetilde{\lambda}_{1a}&-4\upsilon_a  \upsilon_1\widetilde{\lambda}_{1a} & 4 \upsilon_a^2 \lambda_a-\frac{\upsilon_1^2}{2} \lambda_{1ab} & 2 \upsilon_a^2 \lambda_a+\frac{\upsilon_1^2}{2} \lambda_{1ab}\\[2pt]
2\upsilon_a \upsilon_1\widetilde{\lambda}_{1a} & -4\upsilon_a  \upsilon_1\widetilde{\lambda}_{1a}  & 2 \upsilon_a^2 \lambda_a+\frac{\upsilon_1^2}{2} \lambda_{1ab} & 4 \upsilon_a^2 \lambda_a-\frac{\upsilon_1^2}{2} \lambda_{1ab}
\end{pmatrix} \,,
\eea

\bea
M_P^2 \simeq
\begin{pmatrix}
-2 \upsilon_a^2 \lambda_{1ab}+18 \upsilon_1^2 \lambda'_{13} & 3\upsilon_1\lambda'_{13} & \upsilon_a \upsilon_1 \lambda_{1ab} & \upsilon_a \upsilon_1 \lambda_{1ab}\\
3 \upsilon_1^2\lambda'_{13} & \cfrac{\upsilon_1^2}{2}\lambda'_{13} & 0 &0 \\
\upsilon_a \upsilon_1 \lambda_{1ab} & 0 & 4\mu'^2_a-\frac{\upsilon_1^2}{2}\lambda_{1ab} &  -\frac{\upsilon_1^2}{2}\lambda_{1ab}\\
\upsilon_a \upsilon_1 \lambda_{1ab} & 0 & -\frac{\upsilon_1^2}{2}\lambda_{1ab} & 4 \mu'^2_a-\frac{\upsilon_1^2}{2}\lambda_{1ab}\\
\end{pmatrix}\,.
\eea
}
The physical basis is the flavon mass basis 
\beq \label{eq:toyFlavRot}
   \sigma_j = \left(U_S\right)_{ij}\, S_i\hspace{5mm},  
    \hspace{5mm}
    \varphi_i = \left(U_P\right)_{ij}\, P_j,
\eeq
defined where $M_S^2$ and $M_P^2$ are diagonal
\bea \label{eq:toyminD2V}
   M^2_S & \longrightarrow &    
    U_S^T\; {M}^2_S\; U_S = {\rm diag}\left(m^2_{S_1},\, m^2_{S_2},\, m^2_{S_3},\, m^2_{S_4} \right), \\
    M^2_P & \longrightarrow &   
    U_P^T\; {M}^2_P\; U_P = {\rm diag}\left(m^2_{P_1},\, m^2_{P_2},\, m^2_{P_3},\, m^2_{P_4} \right).
\eea
with eigenvalues
{\small
\bea
\label{eq:M2hd}
m_{S_1}^2 &\simeq& 2\upsilon_1^2\left(2\lambda_1 - \lambda_{13}-\cfrac{9}{4}\lambda'_{13}\right) \hspace{9mm},\hspace{2mm}
m_{P_1}^2 \simeq -2\,\upsilon_a^2\left(\lambda_{1ab}+\frac{\upsilon_1^2}{2\upsilon_a^2}(\lambda_{1ab}-18\lambda'_{13})\right)\,,\\
m_{S_2}^2 &\simeq& 2\,\upsilon_1^2\left(2\widetilde{\lambda}_1 - \cfrac{5}{2} \cfrac{\widetilde{\lambda}^2_{1a}}{\widetilde{\lambda}_{a}}\right)\hspace{1.9cm},\hspace{2mm}
m_{P_2}^2 \simeq \upsilon_1^2 \cfrac{\lambda'_{13}}{2}\,,\\
m_{S_3}^2 &\simeq& 2\upsilon_a^2\,\left(2\lambda_a-\lambda_{ab}-\frac{\upsilon_1^2}{2\upsilon_a^2}\,\lambda_{1ab}\right)\hspace{4mm},\hspace{2mm}
m_{P_{3,4}}^2 \simeq 4\,{\mu'_a}^2\,,\\
m_{S_4}^2 &\simeq& 2 \upsilon_a^2\,\left( 2\lambda_{a}+\lambda_{ab}+10\,\frac{\upsilon_1^2}{\upsilon_a^2} \cfrac{\widetilde{\lambda}_{1a}^2}{\widetilde{\lambda}_a}\right)\,.
\eea}
where we have expressed $\mu^2_{1,a}$ with their value at the minimum using \eqref{eq:v1vamin}. The diagonalization matrices, $U_S$, $U_P$, at $\mathcal{O}(\upsilon_1/\upsilon_a)$ can be written as
{\small
\bea
\label{eq:USP}
U_S=
\begin{pmatrix}
\cfrac{2}{\sqrt{5}} & -\cfrac{1}{\sqrt{5}} & 0 & \cfrac{\sqrt{2}\upsilon_1}{\upsilon_a}\cfrac{\widetilde{\lambda}_{1a}}{\widetilde{\lambda}_a}\\
\cfrac{1}{\sqrt{5}} & \cfrac{2}{\sqrt{5}} & 0 & \cfrac{-2\sqrt{2}\upsilon_1}{\upsilon_a}\cfrac{\widetilde{\lambda}_{1a}}{\widetilde{\lambda}_a}\\
0& \cfrac{5\sqrt{2}\upsilon_1}{\upsilon_a}\cfrac{\widetilde{\lambda}_{1a}}{\widetilde{\lambda}_a} &- \cfrac{1}{\sqrt{2}} & \cfrac{1}{\sqrt{2}}\\
0& \cfrac{5\sqrt{2}\upsilon_1}{\upsilon_a}\cfrac{\widetilde{\lambda}_{1a}}{\widetilde{\lambda}_a} & \cfrac{1}{\sqrt{2}} & \cfrac{1}{\sqrt{2}}
\end{pmatrix}\hspace{5mm},\hspace{5mm}
U_P=
\begin{pmatrix}
-1 & 0 & \cfrac{\upsilon_1}{\sqrt{2}\upsilon_a} & 0\\
0 & 1 &0 &0\\
\cfrac{\upsilon_1}{\sqrt{2}\upsilon_a} & 0 & \cfrac{1}{\sqrt{2}} & -\cfrac{1}{\sqrt{2}}\\
\cfrac{\upsilon_1}{\sqrt{2}\upsilon_a} & 0 & \cfrac{1}{\sqrt{2}} & \cfrac{1}{\sqrt{2}}\\
\end{pmatrix}\,.
\eea
}
The computation of the radiative diagram and the contribution to the anomalous magnetic moment in the flavon mass basis are given by
\bea 
\label{eq:toyRADmass}
    & m_\e^{\rm RAD} =  2\, \cfrac{g_\e^4}{16\pi^2}\, \cfrac{\upsilon_H}{\sqrt{2}}\, 
                  \varepsilon_1 \; \Delta I_m \hspace{5mm},\hspace{5mm}
    m_\mu^{\rm RAD}  =   \cfrac{g_\mu^3}{16\pi^2}\, \cfrac{\upsilon_H}{\sqrt{2}}\; 
                 \Delta I_m,\\
\label{eq:toyg2}
&\Delta a_e  =   2\cfrac{ g_\e^4}{8\pi^2}\, \cfrac{\upsilon_H}{\sqrt{2}}\,\cfrac{m_\e}{\mX^2}\, 
                 \varepsilon_1 \; \Delta I_{\Delta a}  \hspace{5mm},\hspace{5mm}
\Delta a_\mu  = \cfrac{g_\mu^3}{8\pi^2}\, 
                 \cfrac{\upsilon_H}{\sqrt{2}}\,\cfrac{m_\mu}{\mX^2}\, \Delta I_{\Delta a}.
\eea
where we have defined
\bea
\Delta I = \sum_{i=1}^{4}\; \left[(U_{S})_{1,i}^2\, I(x_{S_i}^2) \;-\; (U_{P})_{1,i}^2\, I(x_{P_i}^2)\,\right]\,.
\eea
In the case $M_\chi, \mu_{\phi,1} \gg m_\ell$, the loop functions 
are
\bea
I_m(x_\phi) = \frac{1-x_\phi^2(1-2\log x_\phi) }{1-x_\phi^2} \quad , \quad I_{\Delta a}(x_\phi) = \frac{1-4x_\phi^2+x_\phi^4(3-4\log{x_\phi})}{2(1-x_\phi^2)^3}\,.
\eea
From Eq.(\ref{eq:USP}) we see that, up to order $\mathcal{O}(\upsilon_1/\upsilon_a)$, we have  $(U_{S})_{1,i}=(2/\sqrt{5},-1/\sqrt{5},0,0)$ and $(U_{P})_{1,i}=(-1,0,0,0)$. Therefore, as already mentioned, in the calculation of $m^{\rm RAD}_\ell$ and $\Delta a_\ell$ only $S_{1,2}$ and $P_{1}$  play a significant role.
The Eqs.(\ref{eq:toyRADmass}) and (\ref{eq:toyg2}) are very well approximated by the Mass Insertion Approximations of Eqs.(\ref{eq:mmutot}-\ref{eq:toy(g-2)e}).

\acknowledgments

The authors thank A. Santamaria for useful discusions. AM and OV were supported  by Spanish and European funds under MICIU Grant FPA2017-84543-P and by the ``Centro de Excelencia Severo Ochoa'' programme under grant SEV-2014-0398. OV acknowledges partial support from the ``Generalitat Valenciana'' grant PROMETEO2017-033.
AM acknowledges support from La-Caixa-Severo Ochoa scholarship.

\bibliographystyle{JHEP}
\bibliography{biblio}

\end{document}